\documentclass[tradit, longauth]{aa}

\usepackage{dblfloatfix}

\usepackage{amsmath}
\usepackage{natbib, twoopt}

\usepackage{tablefootnote}

\usepackage{graphicx}
\usepackage{color}
\usepackage{url}

\usepackage{txfonts}

\makeatletter
\newcommand\tinyv{\@setfontsize\tinyv{4pt}{6}}
\renewcommand*{\@fnsymbol}[1]{\ifcase#1\or*\or$\dagger$\or$\ddagger$\or**\or$\dagger\dagger$\or$\ddagger\ddagger$\fi}
\makeatother

\definecolor{blond}{rgb}{0.98, 0.94, 0.75}
\definecolor{orange}{rgb}{0.996, 0.847, 0.694}
\definecolor{rose}{rgb}{1.0,0.894,0.882}
\definecolor{violet}{rgb}{0.86, 0.82, 1}
\definecolor{azure}{rgb}{0.80, 1.0, 1.0}

\definecolor{lavender}{rgb}{0.96, 0.73, 1.0}

\def\aj{AJ}                   

\def\araa{ARA\&A}             
\def\apj{ApJ}                 
\def\apjl{ApJ}                
\def\apjs{ApJS}               
\def\aap{A\&A}                
\def\aapr{A\&A~Rev.}          
\def\mnras{MNRAS}             
\def\prl{Phys.~Rev.~Lett.}    
\def\pasj{PASJ}               
\def\prd{Phys.~Rev.~D.}
\def\nat{Nature}              

\makeatletter
\renewcommand*{\@fnsymbol}[1]{\ifcase#1\or*\or$\dagger$\or$\ddagger$\or**\or$\dagger\dagger$\or$\ddagger\ddagger$\fi}
\makeatother

\begin{document}


\titlerunning{A shock-compressed magnetic field in Vela Jr.}
\authorrunning{IXPE team}

\title{Evidence for a shock-compressed magnetic field in the northwestern rim of Vela Jr. from X-ray polarimetry}

\author{
     Dmitry~A.~Prokhorov\inst{\ref{D7}, \ref{D8}}\thanks{E-mail: dmitry.prokhorov@uni-wuerzburg.de}\and
     Yi-Jung~Yang\inst{\ref{D6}, \ref{D4}}\and
     Riccardo~Ferrazzoli\inst{\ref{I8}}\and
     Jacco~Vink\inst{\ref{IS30}}\and
     Patrick~Slane\inst{\ref{I6}}\and
     Enrico~Costa\inst{\ref{I8}} \and
     Stefano~Silvestri\inst{\ref{IS4}}\and
     Ping~Zhou\inst{\ref{D2}}\and
     Niccol\`{o}~Bucciantini\inst{\ref{IS8},\ref{IS9},\ref{IS10}} \and
     Alessandro~Di Marco\inst{\ref{I8}}\and
     Martin~C.~Weisskopf\inst{\ref{I16}}\and
    Luca~Baldini\inst{\ref{IS4},\ref{IS5}} \and
    Victor~Doroshenko\inst{\ref{IS13}}\and
    Steven~R.~Ehlert\inst{\ref{I16}}\and
    Jeremy~Heyl\inst{\ref{IS16}}\and
    Philip~Kaaret\inst{\ref{I16}} \and
    Dawoon~E.~Kim\inst{\ref{I8},\ref{I21},\ref{D1}}\and
    Fr\'{e}d\'{e}ric~Marin\inst{\ref{I10}}\and
    Tsunefumi~Mizuno\inst{\ref{IS25}}\and
    Chi-Yung~Ng\inst{\ref{IS27}}\and
    Melissa~Pesce-Rollins\inst{\ref{IS4}}\and
    Carmelo~Sgr\`{o}\inst{\ref{IS4}}\and
    Paolo~Soffitta\inst{\ref{I8}}  \and
    Douglas~A.~Swartz\inst{\ref{IS11}}\and
    Toru~Tamagawa\inst{\ref{IS14}}\and
    Fei~Xie\inst{\ref{IS31},\ref{I8}}\and
    Iv\'an~Agudo\inst{\ref{IS1}} \and
    Lucio~A.~Antonelli\inst{\ref{IS2},\ref{IS3}} \and
    Matteo~Bachetti\inst{\ref{I19}} \and
    Wayne~H.~Baumgartner\inst{\ref{I16}} \and
    Ronaldo~Bellazzini\inst{\ref{IS4}}\and
    Stefano~Bianchi\inst{\ref{I1}} \and
    Stephen~D.~Bongiorno\inst{\ref{I16}} \and
    Raffaella~Bonino\inst{\ref{IS6},\ref{IS7}}  \and
    Alessandro~Brez\inst{\ref{IS4}}\and
    Fiamma~Capitanio\inst{\ref{I8}} \and
    Simone~Castellano\inst{\ref{IS4}} \and
    Elisabetta~Cavazzuti\inst{\ref{I18}} \and
    Chien-Ting~Chen\inst{\ref{IS11}} \and
    Stefano~Ciprini\inst{\ref{I22},\ref{IS3}} \and
    Alessandra ~De Rosa\inst{\ref{I8}} \and
    Ettore~Del Monte\inst{\ref{I8}}\and
    Laura~Di Gesu\inst{\ref{I18}}\and
    Niccol\`{o}~Di Lalla\inst{\ref{IS12}}\and
    Immacolata~Donnarumma\inst{\ref{I18}}\and
    Michal~Dov\v{c}iak\inst{\ref{I2}}\and
    Teruaki~Enoto\inst{\ref{IS14}}\and
    Yuri~Evangelista\inst{\ref{I8}}\and
    Sergio~Fabiani\inst{\ref{I8}}\and
    Javier~A.~Garc\'{i}a\inst{\ref{I14}}  \and
    Shuichi~Gunji\inst{\ref{I15}}  \and
    Wataru~Iwakiri\inst{\ref{IS17}}\and
    Svetlana~G.~Jorstad\inst{\ref{IS18},\ref{IS19}}\and
    Vladimir~Karas\inst{\ref{I2}}\and
    Fabian~Kislat\inst{\ref{IC}}\and
    Takao~Kitaguchi\inst{\ref{IS14}}\and
    Jeffery~J.~Kolodziejczak\inst{\ref{I16}}\and
    Henric~Krawczynski\inst{\ref{I5}} \and
    Fabio~La Monaca\inst{\ref{I8},\ref{I21},\ref{D1} }\and
    Luca~Latronico\inst{\ref{IS6}}\and
    Ioannis~Liodakis\inst{\ref{I16}}\and
    Simone~Maldera\inst{\ref{IS6}}\and
    Alberto~Manfreda\inst{\ref{IS21}}\and
    Andrea~Marinucci\inst{\ref{I18}}  \and
    Alan~P.~Marscher\inst{\ref{IS18}}\and
    Herman~L.~Marshall\inst{\ref{IS23}}\and
    Francesco~Massaro\inst{\ref{IS6},\ref{IS7}}\and
    Giorgio~Matt\inst{\ref{I1}} \and
    Ikuyuki~Mitsuishi\inst{\ref{IS24}}\and
    Fabio~Muleri\inst{\ref{I8}}\and
    Michela~Negro\inst{\ref{IS26}}\and
    Stephen~L.~O'Dell\inst{\ref{I16}}\and
    Nicola~Omodei\inst{\ref{IS12}}\and
    Chiara~Oppedisano\inst{\ref{IS6}}\and
    Alessandro~Papitto\inst{\ref{IS2}}\and
    George~G.~Pavlov\inst{\ref{IS28}}\and
    Abel~L.~Peirson\inst{\ref{IS12}}\and
    Matteo~Perri\inst{\ref{IS3},\ref{IS2}}\and
    Pierre-Olivier~Petrucci\inst{\ref{I20}} \and
    Maura~Pilia\inst{\ref{I19}}\and
    Andrea~Possenti\inst{\ref{I19}}\and
    Juri~Poutanen\inst{\ref{I11}}  \and
    Simonetta~Puccetti\inst{\ref{IS3}}\and
    Brian~D.~Ramsey\inst{\ref{I16}}\and
    John~Rankin\inst{\ref{I8}}\and
    Ajay~Ratheesh\inst{\ref{I8}}  \and
    Oliver~J.~Roberts\inst{\ref{IS11}}\and
    Roger~W.~Romani\inst{\ref{IS12}}\and
    Gloria~Spandre\inst{\ref{IS4}}\and
    Fabrizio~Tavecchio\inst{\ref{IS29}}\and
    Roberto~Taverna\inst{\ref{I12}}  \and
    Yuzuru~Tawara\inst{\ref{IS24}}\and
    Allyn~F.~Tennant\inst{\ref{I16}}\and
    Nicholas~E.~Thomas\inst{\ref{I16}}\and
    Francesco~Tombesi\inst{\ref{I21},\ref{I22}} \and
    Alessio~Trois\inst{\ref{I19}}\and
    Sergey~S.~Tsygankov\inst{\ref{I11}}\and
    Roberto~Turolla\inst{\ref{I12},\ref{I24}}\and
    Kinwah~Wu\inst{\ref{I24}}\and
    Silvia~Zane\inst{\ref{I24}}
}

\institute{
    Fakult{\"a}t f{\"u}r Physik und Astronomie,  Julius-Maximilians-Universit{\"a}t W{\"u}rzburg, Emil-Fischer-Str. 31, 97074 W{\"u}rzburg, Germany \label{D7} \and
    Institute of Physics of Academia Sinica, No. 128, Section 2, Academia Road, Nangang District, Taipei City, Taiwan \label{D8}\and
    Graduate Institute of Astronomy, National Central University, 300 Zhongda Road, Zhongli, Taoyuan 32001, Taiwan \label{D6} \and
    Laboratory for Space Research, The University of Hong Kong, Cyberport 4, Hong Kong  \label{D4} \and
    INAF Istituto di Astrofisica e Planetologia Spaziali, Via del Fosso del Cavaliere 100, 00133 Roma, Italy \label{I8}\and
    Anton Pannekoek Institute for Astronomy \& GRAPPA, University of Amsterdam, Science Park 904, 1098 XH Amsterdam, The Netherlands\label{IS30}\and
    Center for Astrophysics, Harvard \& Smithsonian, 60 Garden St, Cambridge, MA 02138, USA \label{I6}\and
    Istituto Nazionale di Fisica Nucleare, Sezione di Pisa, Largo B. Pontecorvo 3, 56127 Pisa, Italy\label{IS4}\and
    School of Astronomy and Space Science, Nanjing University, Nanjing 210023, People’s Republic of China \label{D2} \and
    INAF Osservatorio Astrofisico di Arcetri, Largo Enrico Fermi 5, 50125 Firenze, Italy\label{IS8}\and
    Dipartimento di Fisica e Astronomia, Universit\`{a} degli Studi di Firenze, Via Sansone 1, 50019 Sesto Fiorentino (FI), Italy\label{IS9}\and
    Istituto Nazionale di Fisica Nucleare, Sezione di Firenze, Via Sansone 1, 50019 Sesto Fiorentino (FI), Italy\label{IS10}\and
    NASA Marshall Space Flight Center, Huntsville, AL 35812, USA \label{I16}\and
    Dipartimento di Fisica, Universit\`{a} di Pisa, Largo B. Pontecorvo 3, 56127 Pisa, Italy\label{IS5}\and
    Institut f\"{u}r Astronomie und Astrophysik, Universität Tübingen, Sand 1, 72076 T\"{u}bingen, Germany\label{IS13}\and
    University of British Columbia, Vancouver, BC V6T 1Z4, Canada\label{IS16}\and
    Dipartimento di Fisica, Universit\`{a} degli Studi di Roma ``Tor Vergata'', Via della Ricerca Scientifica 1, 00133 Roma, Italy \label{I21}\and
    Dipartimento di Fisica, Universit\`{a} degli Studi di Roma ``La Sapienza'', Piazzale Aldo Moro 5, 00185 Roma, Italy \label{D1}\and
    Universit\'{e} de Strasbourg, CNRS, Observatoire Astronomique de Strasbourg, UMR 7550, 67000 Strasbourg, France\label{I10}\and
    Hiroshima Astrophysical Science Center, Hiroshima University, 1-3-1 Kagamiyama, Higashi-Hiroshima, Hiroshima 739-8526, Japan\label{IS25}\and
    Department of Physics, University of Hong Kong, Pokfulam, Hong Kong\label{IS27}\and
    Science and Technology Institute, Universities Space Research Association, Huntsville, AL 35805, USA\label{IS11}\and
    RIKEN Cluster for Pioneering Research, 2-1 Hirosawa, Wako, Saitama 351-0198, Japan\label{IS14}\and
    Guangxi Key Laboratory for Relativistic Astrophysics, School of Physical Science and Technology, Guangxi University, Nanning 530004, China\label{IS31}\and
    Instituto de Astrof\'{i}sica de Andaluc\'{i}a -- CSIC, Glorieta de la Astronom\'{i}a s/n, 18008 Granada, Spain\label{IS1}\and
    INAF Osservatorio Astronomico di Roma, Via Frascati 33, 00040 Monte Porzio Catone (RM), Italy\label{IS2}\and
    Space Science Data Center, Agenzia Spaziale Italiana, Via del Politecnico snc, 00133 Roma, Italy\label{IS3}\and
    INAF-Osservatorio Astronomico di Cagliari, via della Scienza 5, I-09047 Selargius (CA), Italy \label{I19}\and
    Dipartimento di Matematica e Fisica, Università degli Studi Roma Tre, Via della Vasca Navale 84, 00146 Roma, Italy \label{I1} \and
    Istituto Nazionale di Fisica Nucleare, Sezione di Torino, Via Pietro Giuria 1, 10125 Torino, Italy\label{IS6}\and
    Dipartimento di Fisica, Universit\`{a} degli Studi di Torino, Via Pietro Giuria 1, 10125 Torino, Italy\label{IS7}\and
    Agenzia Spaziale Italiana, Via del Politecnico snc, 00133 Roma, Italy \label{I18}\and
    Istituto Nazionale di Fisica Nucleare, Sezione di Roma ``Tor Vergata'', Via della Ricerca Scientifica 1, 00133 Roma, Italy \label{I22}\and
    Department of Physics and Kavli Institute for Particle Astrophysics and Cosmology, Stanford University, Stanford, California 94305, USA\label{IS12}\and
    Astronomical Institute of the Czech Academy of Sciences, Bo\v{c}n\'{i} II 1401/1, 14100 Praha 4, Czech Republic \label{I2} \and
    California Institute of Technology, Pasadena, CA 91125, USA \label{I14}\and
    Yamagata University,1-4-12 Kojirakawa-machi, Yamagata-shi 990-8560, Japan \label{I15}\and
    International Center for Hadron Astrophysics, Chiba University, Chiba 263-8522, Japan\label{IS17}\and
    Institute for Astrophysical Research, Boston University, 725 Commonwealth Avenue, Boston, MA 02215, USA\label{IS18}\and
    Department of Astrophysics, St. Petersburg State University, Universitetsky pr. 28, Petrodvoretz, 198504 St. Petersburg, Russia\label{IS19}\and
    Department of Physics and Astronomy and Space Science Center, University of New Hampshire, Durham, NH 03824, USA\label{IC}\and
    Physics Department, McDonnell Center for the Space Sciences, and905
Center for Quantum Leaps, Washington University in St. Louis, St.906
Louis, MO 63130, USA\label{I5}\and
Istituto Nazionale di Fisica Nucleare, Sezione di Napoli, Strada Comunale Cinthia, 80126 Napoli, Italy\label{IS21}\and
    MIT Kavli Institute for Astrophysics and Space Research, Massachusetts Institute of Technology, 77 Massachusetts Avenue, Cambridge, MA 02139, USA\label{IS23}\and
        Graduate School of Science, Division of Particle and Astrophysical Science, Nagoya University, Furo-cho, Chikusa-ku, Nagoya, Aichi 464-8602, Japan\label{IS24}\and
        Department of Physics and Astronomy, Louisiana State University, Baton Rouge, LA 70803, USA\label{IS26}\and
        Department of Astronomy and Astrophysics, Pennsylvania State University, University Park, PA 16801, USA\label{IS28}\and
            Univ. Grenoble Alpes, CNRS, IPAG, 38000 Grenoble, France \label{I20}\and
            Department of Physics and Astronomy, 20014 University of Turku, Finland\label{I11}\and
            INAF Osservatorio Astronomico di Brera, via E. Bianchi 46, 23807 Merate (LC), Italy\label{IS29}\and
            Dipartimento di Fisica e Astronomia, Universit\`{a} degli Studi di Padova, Via Marzolo 8, 35131 Padova, Italy \label{I12}\and
    Mullard Space Science Laboratory, University College London, Holmbury St Mary, Dorking, Surrey RH5 6NT, UK \label{I24}
}


\abstract{Synchrotron X-ray emission has been detected from nearly a dozen young supernova remnants (SNRs). X-rays of synchrotron origin exhibit linear polarization in a regular, non-randomly oriented magnetic field. The significant polarized X-ray emission from four such SNRs has already been reported on the basis of observations with the Imaging X-ray Polarimetry Explorer (IXPE). The magnetic-field structure as derived from IXPE observations is radial for Cassiopeia A, Tycho's SNR, and SN 1006, and tangential for RX J1713.7-3946. The latter together with the recent detection of a tangential magnetic field in SNR 1E 0102.2-7219 by the Australia Telescope Compact Array in the radio band shows that tangential magnetic fields can also be present in young SNRs. Thus, the dichotomy in polarization between young and middle-aged SNRs (radial magnetic fields in young SNRs, but tangential magnetic fields in middle-aged SNRs), previously noticed in the radio band, deserves additional attention. The present analysis of IXPE observations determines, for the first time, a magnetic-field structure in the northwestern rim of Vela Jr, also known as RX J0852.0$-$4622, and provides a new example of a young SNR with a tangential magnetic field. }


\maketitle

\makeatletter
\renewcommand*{\@fnsymbol}[1]{\ifcase#1\@arabic{#1}\fi}
\makeatother

\renewcommand{\labelitemi}{$\bullet$}

\section{Introduction}
\label{Sect1}

Shocks in supernova remnants (SNRs) transform part of the bulk kinetic energy of supernova ejecta into energy of nonthermal particles \citep[][]{Krymskii1977, Bell1978, Blandford1978}. It is believed that this process is due to diffusive shock acceleration also known as first-order Fermi acceleration, which produces a power-law spectrum of nonthermal particles. It is widely accepted that particle acceleration at shocks of SNRs provides a theoretical explanation for the majority of cosmic rays with multi-teraelectronvolt (TeV) energies pummeling Earth's atmosphere \citep[for a review,][]{Berezinskii1990}.
SNRs' shocks are capable of accelerating both nuclei and electrons, but the rates at which particles are accelerated may depend on various factors. If accelerated electrons gyrating in SNRs' magnetic fields are sufficiently energetic, they emit synchrotron X-rays \citep[for a review, see][]{Reynolds2008}. The strength of magnetic fields measured near shocks is significantly greater than expected for shock compression of an interstellar magnetic field (IMF) \citep[for a review, see][]{Helder2012}, making the synchrotron process even more important in characterizing the physical conditions at shocks. The explanation of their strength is in magnetic-field amplification upstream of the shock \citep[e.g.,][]{Bell2001, Bell2004}.

X-ray observations provide crucial information to understand the physical conditions at SNRs' shock fronts \citep[for a review,][]{Vink2012}.
Observations with the Advanced Satellite for Cosmology and Astrophysics (ASCA) showed, for the first time, that the X-ray emissions from some SNRs are of synchrotron origin \citep[][]{Koyama1995, Koyama1997}.
Nowadays, almost a dozen SNRs are known emitters of synchrotron X-rays \citep[see][and the reference therein]{Helder2012}. Electrons with multi-TeV energies are responsible for production of these nonthermal X-rays. The synchrotron emission heavily dominates the X-ray spectra of a handful of these SNRs -- namely, SNR G1.9+0.3, SNR G330.2+1.0, RX J1713.7-3946, Vela Jr., and SNR G353.6-0.7.
X-ray imaging of SNRs with a sub-arcsec angular resolution available with Chandra   allowed proper motions of shock waves to be measured \citep[e.g.,][]{DeLaney2003}. High-resolution X-ray imaging, furthermore, provided a useful method to measure magnetic-field strengths in post-shock regions \citep[][]{Vink2003, Voelk2005, Parizot2006}. This method takes into account that the lifetime of multi-TeV electrons due to synchrotron losses is much shorter than the age of an SNR; it relies on the assumption that the measured widths of X-ray filaments do not depend on magnetic field decay. The magnetic-field strengths are 120-250 $\mu$G for Cassiopeia A (Cas A), Tycho's SNR, and Kepler's SNR, and 30-80 $\mu$G for RX J1713.7-3946, RCW 86, and Vela Jr \citep[see][]{Helder2012}.

In the magnetic fields inferred in SNRs, electrons with gigaelectronvolt (GeV) energies emit via the synchrotron process in the radio band. However, these radio-wave emitting electrons have much longer lifetimes than those emitting X-rays. Polarimetric observations at radio frequencies established a dichotomy between young ($\lesssim$ 1000-year-old) SNRs (Cas A, Tycho's SNR, Kepler's SNR, and SN 1006) and middle-aged ($\simeq$ 10,000-year-old) SNRs \citep[for a review, see][]{Dubner2015}. In fact, the former ones have a tangential polarization (the orientation of electric vectors), while the latter ones have a radial polarization. Due to the shock compression, the tangential  component of a magnetic field in a post-shock region (that is perpendicular to the shock normal) increases, but the radial component does not change. This would always lead to a radial polarization. However, plasma instabilities may stretch a magnetic field in the direction of bulk motion \citep[e.g.,][]{Gull1973, JN962, JN961, Inoue2013}, resulting in a tangential polarization. The tangential polarization may alternatively be created due to a selection effect, which favors particle acceleration for the quasi-parallel part of the shock \citep[][]{West2017}.

The Imaging X-ray Polarimetry Explorer \citep[IXPE;][]{Weisskopf2022} allows the measurement of polarized emission from SNRs with synchrotron X-ray spectral components. The first three SNRs observed with IXPE were Cas A \citep[][]{VinkIXPE}, Tycho SNR \citep[][]{FerrazzoliIXPE}, and the northeastern limb of SN 1006 \citep[][]{ZhouIXPE}.
The tangential X-ray polarization pattern \citep[reported by][]{VinkIXPE, FerrazzoliIXPE, ZhouIXPE} is compatible with the corresponding polarization pattern measured in the radio band \citep[see, e.g.,][for Cas A, Tycho's SNR, and the northeastern limb of SN 1006, respectively]{Rosenberg1970, Kundu1971, Reynoso2013}. The X-ray polarization indicated that plasma instabilities in these young SNRs act to produce radial magnetic fields very close to the shock fronts given the short lifetimes of X-ray-emitting electrons. To better understand the dichotomy in polarization between young and middle-aged SNRs, IXPE was pointed to the northwestern (NW) region of RX J1713.7-3946. With the age of about 1600 years, this SNR is older than the first three SNRs that were observed. The IXPE observations of RX J1713.7-3946 resulted in a radial polarization \citep[][]{Ferrazzoli2024}. In the radio band, recent measurements with the Australia Telescope Compact Array (ATCA) also revealed a radial polarization in SMC SNR 1E 0102.2-7219 \citep[][]{Alsaberi2024}, which has an age of 1700 years. We note that RX J1713.7-3946 and SNR 1E 0102.2-7219 are much younger than the previously known SNRs with the radial polarization pattern.

Vela Jr. (also known as RX J0852.0-4622 or SNR G266.2-1.2) is a Galactic SNR discovered in the X-ray data of R\"{o}ntgensatellit (ROSAT) at the south-east corner of the known Vela SNR \citep[][]{Aschenbach1998}. Its X-ray spectrum is featureless and well described by a power law \citep[][]{Slane2001}. The detection of Vela Jr. in very-high-energy $\gamma$ rays \citep[][]{Komin2005} provided strong confirmation that it is an SNR in its own right, and not a substructure within the larger Vela SNR. Vela Jr. has an angular diameter of $\simeq2^{\circ}$, with a peak of X-ray emission in the NW rim that is $\simeq5^{\prime}$ in size \citep[][]{Bamba2005, Mayer2023}. The expansion rate of the NW rim of Vela Jr. is about 5 times lower than that in Cas A, suggesting that its age is between 1700 years and 4300 years \citep[e.g.,][]{Katsuda2008}.
The presence of a central compact object, similar to the central compact source of the Cas A, indicates that Vela Jr. was born from a core-collapse supernova \citep[][]{Pavlov2001}. Both the age and the massive progenitor make this SNR similar to RX J1713.7-3946 and 1E 0102.2-7219. Vela Jr. and RX J1713.7-3946 have similarly high effectiveness of a magnetic turbulence in diffusing particles across a shock front. This property is usually characterized by the Bohm factor.
The Bohm factors, $\eta$, measured in the NW rims of RX J1713.7-3946 and Vela Jr. are $1.4\pm0.3$ and $0.7\pm0.5$, respectively, and significantly smaller than those for Cas A, Tycho, and SN 1006 \citep[e.g.,][]{Tsuji2021}. The smallness of the Bohm factors in the NW rims of RX J1713.7-3946 and Vela Jr. suggests that the acceleration proceeds in a regime close to the Bohm limit of $\eta=1$ and that the particles are accelerated most efficiently in these regions. The Bohm factor itself is related to the spectrum of the turbulent magnetic field that scatters particles. The radio emission from Vela Jr. and RX J1713.7-3946 is of lower surface brightness than that of Cas A, Tycho SNR, and SN 1006 \citep[][]{Duncan2000, Lazendic2004}. The radio emission from Vela Jr. has a surface brightness similar to that of the foreground Vela SNR. The NW region of Vela Jr. appears polarized to a level of $\sim$20\% at 2.42 GHz, but this polarized radio emission was entirely attributed to the Vela SNR for good reasons \citep[see][]{Duncan2000}. X-ray polarimetry overcomes these limitations for Vela Jr., similar to how it did previously for RX J1713.7-3946, due to its higher X-ray brightness above 2 keV compared to the Vela SNR.

With the purpose of a further study of X-ray polarization in young SNRs with Bohm factors of $\approx 1$, IXPE performed observations of the X-ray-bright NW rim of Vela Jr. After the discovery of a radial polarization in the NW rim of RX J1713.7-3946, a pressing question was whether Vela Jr. has also a radial polarization due to the mentioned similarities.
This paper reports the magnetic-field structure in the NW rim of Vela Jr. inferred from IXPE data.

\section{Observations}
\label{sec2}

\subsection{Instrument} 

Launched on 2021 December 9, IXPE resides in a low-Earth equatorial orbit.
IXPE contains three grazing-incidence X-ray telescopes, each consisting of a 4-m-focal-length mirror module assembly \citep[][]{Ramsey2022} and a detector unit (DU) hosting a polarization-sensitive gas-pixel detector \citep[][]{Costa2001, Bellazzini2007, Baldini2021}, located at the focus of an X-ray mirror module.
The effective area of the mirror module assemblies, together with the quantum efficiency of DUs, defines the IXPE energy range of 2-8 keV.
The instrument provides an angular resolution of 24$^{\prime\prime}$-30$^{\prime\prime}$ (half-power diameter) and enables imaging X-ray polarimetry of extended sources, such as SNRs. The overlap of the fields of view of the three DUs is circular with a diameter of 12\farcm9, limited by the sensitive area of each detector, 15$\times$15 mm$^2$, and a fiducial area cut of 13.2$\times$13.6 mm$^2$ applied by the IXPE science operations center. The energy resolution of IXPE is $\Delta E\approx 0.5$ keV at 2 keV and scales as the square root of the energy. The DUs record the tracks of photoelectrons, created as a result of X-ray absorption in the dimethyl-ether gas. For polarized X-rays, the photoelectron has an emission direction peaked at that of the electric field of the X-ray and modulated with a cosine square function. IXPE measures the linear polarization of an X-ray source on a statistical basis by studying the azimuthal distribution of photoelectrons. The field of view, the angular resolution, and the energy range of IXPE allow a spatially resolved polarimetric study of synchrotron emission from the X-ray-bright NW rim of Vela Jr.

\subsection{Planning} 

IXPE observations of faint, extended, X-ray synchrotron-emitting SNRs, such as SN 1006 and RX J1713.7-3946, showed that the detections of signals with a polarization degree (PD) of 10-20\%, can require a megasecond (Ms) exposure time.
The simulations with \texttt{ixpeobssim} \citep[version 30.5.0,][]{Baldini2022}, which is a Python-based simulation and analysis framework developed for the IXPE mission, helped us to select the position and the exposure time for IXPE Vela Jr. observations. The \texttt{ixpeobssim.srcmodel.roi.xChandraObservation} class describes the spectral and spatial properties of a source using a Chandra photon list. These particular simulations use the photon list from the Chandra Vela Jr. observations (Observation ID: 3846) that were taken on 2003 January 5-6 and pointing at RA=$132\fdg283$ and Dec=$-45\fdg629$. To account for the residual IXPE instrumental background, the simulations include an additional, isotropic component implemented with the \texttt{xTemplateInstrumentalBkg} class and similar to that used in \citet[][]{FerrazzoliIXPE}. The performed simulations indicate that the polarized X-ray emission from the NW rim of Vela Jr. is detectable above a 5$\sigma$ level with 1 Ms of the IXPE observations, if PD=30\%.
The synchrotron X-ray emission from extended sources can, indeed, be with such a high PD value as, for example, the IXPE observations of the eastern lobe of SS 433 \citep[][]{Kaaret2024} and the X-ray-bright filament, G0.13-0.11 \citep[][]{Churazov2023} have shown.

\subsection{Data taking} 

IXPE observed the NW rim of Vela Jr. during two different epochs. These observation epochs were from 2023 November 24 to 2023 December 6 and from 2023 December 22 to 2024 January 3, corresponding to exposure times per DU of 0.632 Ms and 0.608 Ms, respectively. The data set comprises a total exposure time of 1.24 Ms per DU. IXPE targeted the NW rim of Vela Jr. at RA=$132\fdg240$ and Dec=$-45\fdg650$ through dithering observations with an amplitude of 0\farcm8. The dithering samples the X-ray source over numerous detector pixels and minimizes the impact of pixel-to-pixel variations. These observations entirely covered the X-ray-bright part of the NW rim and, by this, allowed a source region larger than that used in the \texttt{ixpeobssim} simulations. This is due to the fact that the latter is limited by the boundary of one of the Chandra ACIS chips. The larger source region increases the statistical accuracy over that predicted by the simulations.
These IXPE observations also allowed for a background region projected outside Vela Jr. and as large as the source region (see Fig. \ref{plot_stokesI}). The large background region is useful to estimate the background level with a small statistical uncertainty and also to assess the time variations of both the background flux and polarization.

\subsection{Note on alignment} 

The high-resolution Chandra- and XMM-Newton-based studies showed that the shell of Vela Jr. has an expansion proper motion \citep[see][respectively]{Allen2015, Katsuda2008}. There is a discrepancy in the inferred expansion rates of $0.42\pm0.10$ arcsec yr$^{-1}$ \citep[][]{Allen2015} and $0.84\pm0.23$ arcsec yr$^{-1}$ \citep[][]{Katsuda2008}. 
Thus, the expansion may cause a motion of $8^{\prime\prime}$ or $16^{\prime\prime}$ for the shell in a baseline of 20 yr. A comparison of the IXPE observations with the \texttt{ixpeobssim} simulations based on the Chandra 2003 observations is suggestive of a spatial offset of $\sim20^{\prime\prime}$ in the direction of the proper motion. Given that the spatial offset even larger in size was found in the IXPE observations of SN 1006 \citep[][]{ZhouIXPE} and was explained by the difficulty of the boom-bending correction for extended sources \citep[e.g.,][]{Weisskopf2022}, the interpretation of the spatial offset in the IXPE Vela Jr. observations requires further investigation of systematic effects and beyond the scope of this paper. On the other hand, the alignment between the IXPE X-ray maps from the three DUs during the two Vela Jr. observation epochs was precise.

\section{Data reduction}

Instrumental background induced by charged particles is present in IXPE data, but can be distinguished from X-rays in a probabilistic way\footnote{\url{https://github.com/aledimarco/IXPE-background}}  \citep[e.g.,][]{diMarco2023}. The difference in morphology of the recorded tracks is a key ingredient for this disentanglement. X-rays produce photoelectrons that are detected in the gas-pixel detectors in the form of ionization tracks. Background particles, arriving from outside or produced in the IXPE satellite's passive structures, induce tracks typically more extended, straight, and with a lower charge density than photoelectrons. The systematic study of selection criteria to separate X-ray-induced tracks from background ones resulted in three rejection filters \citep[][]{diMarco2023}, involving (i) the track size, which is the number of ASIC pixels above the threshold in the largest group of contiguous pixels of the event (the main clusters), (ii) the energy fraction, taking into account the ratio between the energy (charge) collected in the main cluster and the one collected in all the detected clusters, and (iii) the number of border pixels.
In general, these three rejection filters remove $\sim$40\% of the background events, while keeping 99\% of the X-rays. The application of this rejection algorithm to the IXPE data collected from extended X-ray sources, such as the NW rim of Vela Jr., significantly reduces the particle background and allows a more sensitive polarimetric analysis.
In the filtered IXPE data, the residual background in the X-ray-bright source region still dominates over the synchrotron emission from Vela Jr. above 4 keV. To minimize the contamination from the residual instrumental background, the analysis presented in this paper includes events with energies between 2 keV and 4 keV. Since the residual background is uniform within the central 5$^{\prime}$-radius region \citep[][]{diMarco2023}, the X-ray data set used for a polarization study in this paper includes only events from this central region.

\begin{table}[h]
\centering
\caption{Bright solar flares during these IXPE observations.}
\begin{tabular}{|c | c | c| c |}
 \hline
 Identificator & Date & Start (MET) & End (MET) \\
 \hline
 Flare 1 & 2023-11-28 & 217960825 & 217981916 \\
 Flare 2 & 2023-12-24 & 220204733 & 220220045 \\
 Flare 3 & 2023-12-31 & 220804567 & 220837180 \\
 Flare 4 & 2024-01-02 & 220978625 & 220998670 \\
 \hline
\end{tabular}
\label{Table1}
\end{table}

\begin{table*}[!h]
\centering
\caption{Set of cuts for removing events produced by charged particles and during solar flares.}
\begin{tabular}{|c | c |}
 \hline
 Parameter & Expression \\
 \hline
 Number of pixels & $>70 + 30 \times E/(1 \mathrm{keV})$  \\
 
 Energy fraction & $<0.8\times\left(1-\exp\left(-(E/(1 \mathrm{keV})+0.25)/1.1\right)\right)+0.004\times E/(1 \mathrm{keV})$ or $>1$\\
 
 Border pixels & $>2$  \\
 
 Photon energy & $>4$ keV  \\
 
 Radial distance & $>5^{\prime}$  \\

 Count rates & $>3\sigma$ above the mean count rate  \\

 Time intervals & Flare 1 (DU 2), Flare 2 (DU 2 \& DU 3), Flare 3 (DU 2 \& DU 3), Flare 4 (DU 2 \& DU 3) \\
\hline
\end{tabular}
\tablefoot{
The polarization analysis does not use the events satisfying either of these expressions.}
\label{Table2}
\end{table*}

The rejection algorithm by \citet[][]{diMarco2023} also removes events due to charged particles associated with coronal mass ejections from the Sun, when applied to this IXPE data set. The rate of charged particles may also be enhanced during time intervals when IXPE crosses the boundaries of the South Atlantic Anomaly. Apart from the backgrounds of charged particles, one needs to take care of background X-rays produced by bright solar flares. These charged-particle and X-ray backgrounds appear as short-term increases in counting rate. In the previous papers, such as \citet[][]{ZhouIXPE} and \citet[][]{Ferrazzoli2024}, the authors examined the distribution of count rates and removed the time intervals showing count rates higher than $3\sigma$ above the mean rate. The procedure that is adopted from \citet[][]{Ferrazzoli2024} allowed us to remove the intervals of a temporarily high background from the Vela Jr. data set. The examination of the cleaned data showed that some short-term increases remain. Most of these increases are around time intervals when bright solar flares occurred. Table \ref{Table1} shows the list of four bright solar flares occurred during these IXPE observations. The brightest of these four solar flares occurred on 2023 December 31. This X5-class flare was the most powerful solar flare in six years. Given that DU 2 and DU 3 are more exposed to the Sun than DU 1, these two DUs are the most affected by bright solar flares. The removal of events during the time intervals corresponding to the four solar flares from the data of DU 2 and corresponding the flares 2-4 from the data of DU 3 eliminates the remaining short-term increases. This step shortened the data sets of DU 2 and DU 3 by 60 ks and 45 ks, respectively, while these two procedures shortened the exposure time per DU by about 5\% or less in total. Table \ref{Table2} lists all the cuts applied to the data.

The assumption that the background emission is unpolarized was essential for detections of polarized X-ray emission from faint, extended sources, including the NE region of SN 1006 \citep[][]{ZhouIXPE}, the NW region of RX J1713.7-3946 \cite[][]{Ferrazzoli2024}, the eastern lobe of SS 433 \citep[][]{Kaaret2024}, and the X-ray-bright filament, G0.13-0.11 \citep[][]{Churazov2023}. In these previous studies, it was checked and found that the background emission is, indeed, unpolarized. When approaching the solar maximum, bright solar flares, such as that occurred on 2023 December 31, become more frequent. The IXPE solar panels are fixed perpendicular to the primary axis of the observatory and point within 25$^{\circ}$ from the Sun. The angle between the primary axis of the observatory and the Sun is close to 90$^{\circ}$. When a bright solar flare occurs, solar X-rays enter at $\sim$90$^{\circ}$ in the instrument and this can induce a highly polarized background filling the entire field of view.
With the purpose to check whether background emission during solar flares is polarized, this study of background emission uses the data recorded during the brightest X-ray flare (Flare 3 from Table \ref{Table1}), including the data only from DU 2 and DU 3, but without applying the 3-$\sigma$ clipping procedure. The \texttt{PCUBE} algorithm of the \texttt{xpbin} tool from \texttt{ixpeobssim} allows both extraction of the Stokes parameters of the events collected in a given region and calculation of polarization properties. The region used in this particular study is the central 5$^{\prime}$ region with the exclusion of the X-ray-bright part. The region excluded here serves as the source region in the next Section. The \texttt{PCUBE} analysis of the selected region reveals polarized emission at a high significance level. In the 2-3 keV band, the PD value is as high as $\sim$70\%. This also shows that the exclusion of time intervals corresponding to solar flares is important for a study of polarized signals from faint, extended, X-ray sources. More details on a polarized component during solar flares will be reported in a forthcoming paper. A thorough check of the background emission recorded during the time intervals selected for the present analysis of Vela Jr. showed no evidence of polarization.

Although the Vela SNR is one of the brightest regions in the X-ray sky in the 0.5-1.0 keV band, the contribution of this old SNR to the total emission from the X-ray-bright NW rim of Vela Jr. in the 2-4 keV band is small \citep[see, Figures 3 and 7 in][respectively]{Mayer2023, Camilloni2023}. This subdominant, unpolarized, thermal X-ray component can only slightly reduce the polarization measured from the NW rim of Vela Jr.

\section{Analysis and results}

This section explains the selection of source and background regions and shows the results of \texttt{PCUBE} and spectro-polarimetric analyses of the IXPE Vela Jr. data.

\subsection{Selection of source and background regions}
\label{Sect3p1}

\begin{figure}
\includegraphics[width=0.47\textwidth]{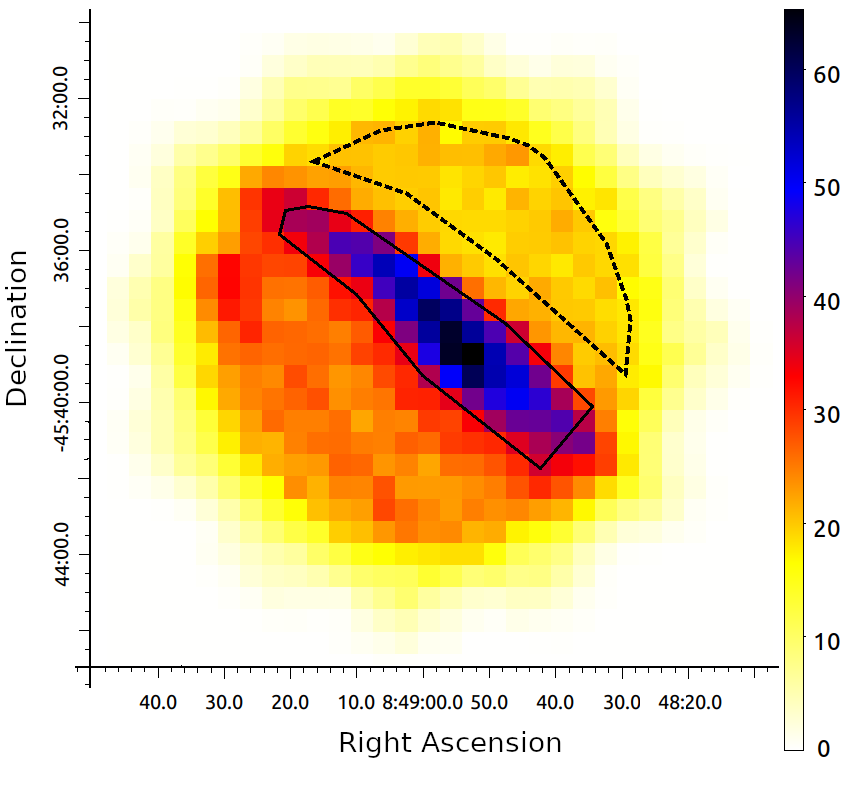}
\caption{Stokes $I$ map (in cm$^{-2}$) with a pixel size of 30$^{\prime\prime}$. The map shows the source and background regions by solid and dashed lines, respectively.}
\label{plot_stokesI}
\end{figure}

The selection of source and background regions is an important step in the analysis. The source region must ideally include the region from where most of the polarized emission comes. The method of selecting the source region introduced and used in this paper assumes a constant PD inside and a uniform polarization angle (PA) across a given region. It only uses the Stokes $I$ map measured by IXPE to estimate the polarized signal.
This method does not rely on the measured polarization properties. Figure \ref{plot_stokesI} shows the Stokes $I$ map produced by the \texttt{PMAPCUBE} algorithm of the \texttt{xpbin} tool. This map has a pixel size of 30$^{\prime\prime}$, which is comparable to the IXPE angular resolution. The bright region in this map corresponds to the X-ray-bright NW rim. The estimator, given by the expression, $I_{\mathrm{corr}}/\sqrt{A}$, where $I_{\mathrm{corr}}$ is the difference between the value of Stokes $I$ for the source region and that estimated for the background region of the same surface area, and $A$ is the surface area covered by the source region. The higher the estimator value, the higher the significance of a polarized signal under the assumption of an identical PD. The source region, shown by a solid line in Figure \ref{plot_stokesI}, provides the highest value of the estimator compared with several different regions probed in this study. The source region has a surface area of $53741$ sqr. arcsec and covers about 60 pixels in Figure \ref{plot_stokesI}. The spatial size of the source region is $\approx2.00\times 0.44$ pc$^2$ at an assumed distance of 750 pc and is large enough for producing steady X-ray emission during the IXPE observations. The number of counts within the source region in the 2-4 keV band is 53869. The estimator for this source region is higher than that for the region used in the simulations on the basis of Chandra data by a factor of 1.4 due to the higher average Stokes $I$ value, although the surface areas of these two regions are almost the same.

Given the telescope pointing position for these IXPE observations, the center of the source region is at the center of the field of view. The source region divides the remaining part of the field of view into two. One half is outside Vela Jr. and the signal from this part is suitable for estimating the total background emission, comprising the instrumental background and the diffuse Galactic background. In contrast, the other half is inside Vela Jr. and, in addition, contains the signal from a minor, fainter X-ray filament seen near the southeastern border. The dashed line in Figure \ref{plot_stokesI} shows the background region selected for this analysis. This region is in the source-free half of the field of view and at a distance of $\approx 75^{\prime\prime}$ from the source region. This distance is sufficient to suppress the contribution from the polarized emission of Vela Jr to the background. The surface area of the background region is $63970$ sqr. arcsec and about 20\% larger than that of the source region. The \texttt{PCUBE} analysis showed that the signal from the background region is unpolarized and with $Q/I=0.014\pm 0.034$ and $U/I=-0.016\pm 0.034$, when one selects the entire time interval, excluding the time intervals listed in Table \ref{Table2}. For consistency, the version of \texttt{ixpeobssim} used for the \texttt{PCUBE} analyses in this paper is the same as that used for the simulations before data collection. The 95\% confidence level upper limit on PD of the background emission is 9.0\%. Thus, the assumption that the background is unpolarized holds for computing the statistical significance of polarized emission from the source region. The background emission contributes $\simeq$38\% of the photons from the source region and the subtraction of background emission is necessary for deriving the polarization properties of emission from the source region.

\subsection{PD distribution}

\begin{table}
\centering
\caption{PD and PA distributions for pixels with more than 3200 counts and a pretrial significance above $2\sigma$.}
\begin{tabular}{| c | c | c | c | c |}
 \hline
 RA  & Dec  & PD$_{\mathrm{obs}}$ & PD$_{\mathrm{corr}}$ & PA$_{\mathrm{obs}}$ \\
(deg) & (deg) & (\%) & (\%) & (deg) \\
 \hline
 132.267 & -45.612 & $19.0\pm 8.3$ & $34.0\pm15.3$ & $142.4\pm12.6$ \\
 132.239 & -45.612 & $34.3\pm 9.7$ & $85.2\pm30.1$ & $151.7\pm8.1$ \\
 132.211 & -45.631 & $32.5\pm 9.0$ & $66.4\pm22.6$ & $151.2\pm8.0$ \\
 132.239 & -45.651 & $19.5\pm 8.4$ & $35.2\pm15.8$ & $122.4\pm12.3$ \\
 132.156 & -45.690 & $24.2\pm 9.7$ & $60.1\pm25.7$ & $151.2\pm11.5$ \\
 \hline
\end{tabular}
\tablefoot{These five pixels are listed in order from east to west, then from north to south in the RA-Dec coordinates.}
\label{Table3}
\end{table}

\begin{figure}
\includegraphics[width=0.5\textwidth]{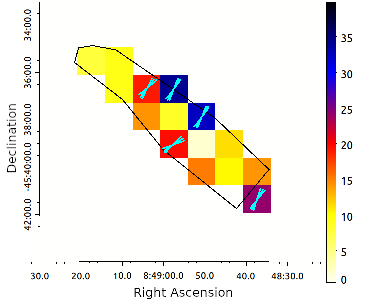}
\caption{PD$_{\mathrm{obs}}$ distribution for pixels with with more than 3200 counts (to mask out pixels with a low X-ray brightness), overlaid with polarization vectors and their 1$\sigma$ errors on PA. The vectors correspond to pixels with a pretrial significance above $2\sigma$ shown in Table \ref{Table3}. }
\label{plot_map}
\end{figure}

The re-binning, produced by means of the \texttt{PMAPCUBE} algorithm of the \texttt{xpbin} tool, to the pixel size of $1^{\prime}$ allows us to select the regions from which signals with PD of $\sim 30\%$ can be measured. The cut, \texttt{COUNTS>3200}, helped us to mask out pixels with low X-ray brightness from which any detection of polarized X-rays is unlikely. After this cut, only 15 pixels remain and all of these pixels correspond to the X-ray-bright NW rim. The examination of the PD map, derived using the \texttt{PMAPCUBE} algorithm, shows that 5 of these 15 pixels correspond to a polarized signal measured with PD/PD$_{\mathrm{err}}>2$. Table \ref{Table3} lists the PD$_{\mathrm{obs}}$ and PD$_{\mathrm{corr}}$ values as observed and as after background subtraction, and the $PA_{\mathrm{obs}}$ values for these 5 pixels. The PA values are measured counterclockwise from north in the equatorial coordinate system. These 5 values of PA are compatible within the uncertainties. The weighted mean of these 5 PA values is $146.6^{\circ}\pm 4.4^{\circ}$, where the weighting factor is the inverse square of the error. Based on the geometry of the shock in the NW rim, a radial polarization corresponds to a PA value of $\simeq 140^{\circ}$. This value is compatible with this weighted mean value. In general, the addition of weakly polarized regions to the more highly polarized region keeps the mean value of PA mostly unchanged but alters that of PD. This makes the PA value derived from the more highly polarized region reasonably representative of the PA value in the entire region. Figure \ref{plot_map} shows the PD distribution for pixels with \texttt{COUNTS>3200}. The solid contour shows the source region selected in Section \ref{Sect3p1}. This region covers most of these 15 pixels. The southernmost of the 15 pixels lies substantially outside the central $5^{\prime}$ region.

\begin{figure}
\includegraphics[width=0.49\textwidth]{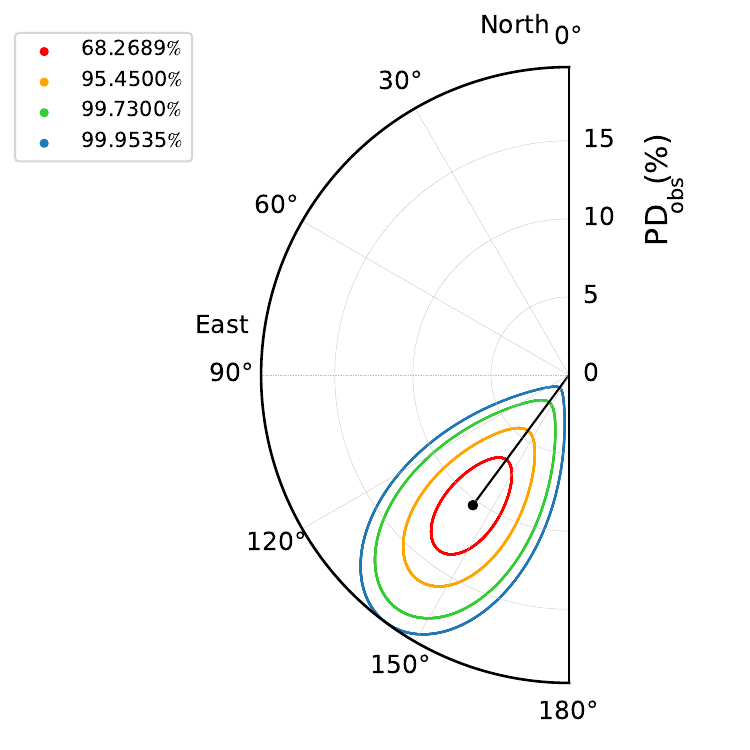}
\caption{Polar plot obtained from \texttt{PCUBE} analysis of the source region. The dot marks the measured PD and PA values. The contours show  $68.27\%$, $95.45\%$, $99.73\%$, and $99.95\%$ confidence levels.}
\label{figure_polarPlot}
\end{figure}

Among these 15 pixels, 5 pixels correspond to higher PD values than the other 10 pixels.
A total of 2 pixels are with a significance of polarized emission at a $2.9-3.0 \sigma$ level based on the \texttt{PCUBE} analysis, but none of these 5 pixels is with a significance above a $3 \sigma$ level. The former 2 pixels have the highest value of PD shown in Figure \ref{plot_map}. The probability to detect such high polarized signals from 2 of the 15 pixels due to a random fluctuation from an unpolarized source is about 0.3\%, that is $15\times14\times(0.0037)^2$, where the p-value of 0.0037 corresponds to a 2.9 $\sigma$ level. This fact hints that the signal from the X-ray-bright NW rim is polarized.

These two pixels lie near the boundary of the X-ray-bright NW rim. For extended sources with sharp edges, false polarization haloes may arise as a consequence of a correlation between the error in reconstructing the X-ray absorption point and the direction of its electric-field vector \citep[see,][for more details]{Bucciantini2023}.
To quantify this effect, known as polarization leakage, on the polarization properties of these pixels, we used the \texttt{leakagelib} code \citep[][]{Dinsmore2024}. The corresponding contributions due to polarization leakage to the measured Stokes Q and U values are only at a level of $\simeq$20\% of the statistical uncertainties on the measured values.

\subsection{Model-independent polarization results}

Figure \ref{figure_polarPlot} shows PD and PA for the entire source region along with the contours indicating four different confidence levels. This polar plot illustrates the results of the \texttt{PCUBE} analysis. Table \ref{Table4} lists the observed PD and PA values and the normalized Stokes $Q$ and $U$ values. The observed PD value is $10.4\%\pm 2.4\%$ and the observed PA value, is $143.6^{\circ}\pm 6.6^{\circ}$. The probability that the observed signal is produced by a random fluctuation from an unpolarized source is $1.6\times10^{-4}$. This is equivalent to a 3.8 $\sigma$ confidence level. The theoretical expectation of a radial polarization in the X-ray-bright NW rim of Vela Jr makes this result more firm. The radial polarization corresponding to the PA value of $\approx 140^{\circ}$ is compatible with the measured value within statistical errors. The joint probability that the observed polarized signal is a random fluctuation from an unpolarized source and has the PA value compatible by mere chance with the radial polarization is $1.6\times10^{-4}\times(2\times6.6/180.0)\simeq1.2\times 10^{-5}$. This provides a strong (4.3 $\sigma$) support to the model resulting in shock-compressed magnetic fields. The NW rim of Vela Jr. is a new example of a young SNR with a radial polarization.

\begin{table}
\centering
\caption{Polarization signal from X-ray-bright NW rim of Vela Jr as observed and as after background subtraction.}
\begin{tabular}{|c | c | c |}
 \hline
 Parameter & Obs. value & Corr. value \\
 \hline
 $Q/I$ & $0.030\pm 0.024$ & $0.060\pm 0.045$ \\
 $U/I$ & $-0.099 \pm 0.024$ & $-0.153\pm 0.045$ \\
 PD & $10.4\%\pm 2.4\%$ & $16.4\%\pm5.2\%$ \\
 PA & $143.6^{\circ}\pm 6.6^{\circ}$ & $145.6^{\circ}\pm9.0^{\circ}$ \\
 \hline
\end{tabular}
\label{Table4}
\end{table}

The contribution of background emission to the signal from the source region is substantial. The \texttt{PCUBE} analysis of the background region provided another set of the values of Stokes $I$, $Q$, and $U$. After scaling these values according to the surface areas, the difference between the corresponding Stokes parameters from the \texttt{PCUBE} analyses of the source and background regions gave the intrinsic values of the Stokes parameters for the synchrotron emission from the source region. Table \ref{Table4} lists the PD and PA values derived from the intrinsic values of the Stokes parameters. This value of PD, $16.4\%\pm 4.5\%$.
This value is similar to PD$=13.0\% \pm 3.5\%$  previously reported for RX J1713.7-3946 \citep[][]{Ferrazzoli2024}. The similarity of the PD values may indicate the similar regularity degrees of magnetic fields near the shock fronts in these SNRs.

\subsection{Smoothed polarization maps}

A common technique in imaging analysis of maps with poor statistics is to smooth images with a  kernel, often a Gaussian kernel. This
improves the statistics at the expense of imaging resolution. It could be argued that rebinning has a similar effect,
but often the results then depend also the centering of the bins, and the result is often less aesthetic. The reason is that smoothing
gives more weight to the central pixel, and gives less weight to pixels that are further out.

In previous papers \citep[e.g.,][]{VinkIXPE}, the polarization signal of  Stokes $Q$ and $U$ maps were in the form of a resulting test statistic maps ($\chi^2_2$ maps).
This concept can also be used for smoothed Stokes $Q$ and $U$ maps.

Formally, smoothing of a map consists of assigning to each pixel a new value, which is the kernel-weighted summation over this pixel, and neighboring pixels.
For example for the Stokes $Q$ map we can write for the smoothed $\tilde{Q}$ map:
$\tilde{Q}_{ij}=\sum_{kl}a_{ijkl}Q_{kl}$,
with $i,j$ the pixel coordinate of the smoothed map, $kl$ the pixels of the input $Q$ map, and $a_{ijkl}=f(k-i,l-j)$ the weights
of the normalized kernel, that is, $\sum_{kl}a_{ijkl}=1$.
The error on the pixel values $\tilde{Q}_{ij}$ can be obtained by quadratic summation
$\sigma(\tilde{Q}_{ij})=\sqrt{ \sum_{kl} [a_{ijkl}\sigma(Q_{kl})]^2}$, with $\sigma$ indicating the statistical error of a quantity.
Since $\sigma(Q_{kl})^2={\rm Var}(Q_{kl})$, we can rewrite this as
\begin{equation}
{\rm Var}(\tilde{Q}_{ij}) =   \sum_{kl}a_{ijkl}^2 {\rm Var}(Q_{kl}).
\end{equation}
So the variance in a smoothed map is obtained by smoothing the variance of the input map with the kernel squared.\footnote{Note that
for a count map, ${\rm Var}(N)=N$, based on Poissonian statistics. So the variance map in that case is the input map smoothed with the kernel squared.}
The gain in signal to noise obtained by smoothing is due to the fact that $\sum_{kl} a^2_{ijkl}<1$; in other words, one divides the images with a variance map with smaller values than the original variance map.

\begin{figure}
  \centerline{
    \includegraphics[width=0.45\textwidth]{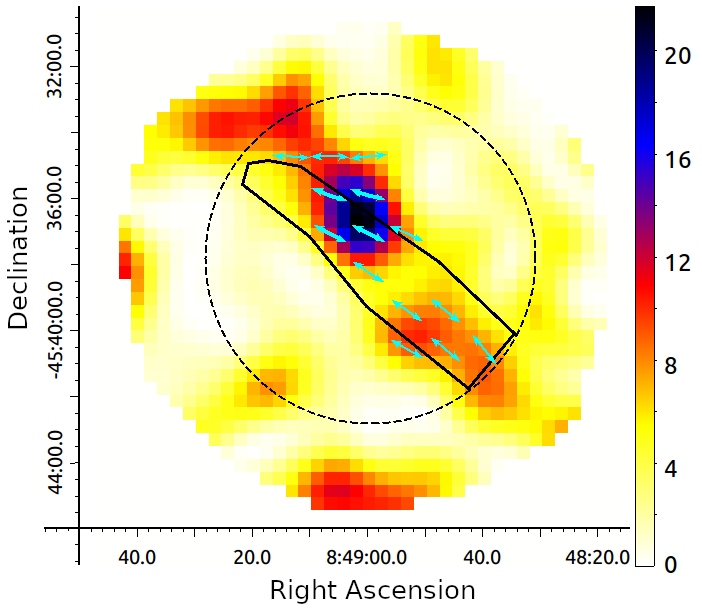}
    }
  \caption{\label{fig:smoothed}
    $\chi^2_2$ map resulting from smoothing the Stokes $I$, $Q$, and $U$ and their variance map with a Gaussian kernel of width $\sigma=58^{\prime\prime}.3$, corresponding to 2.5 pixels.
    The arrows indicate the magnetic-field vectors for pixels with polarization significances $>2\sigma$ within a 5$^{\prime}$ radius circle.
    For regions with $\geq 3\sigma, \chi^2_2\geq 11.8$, the vectors are thicker.
  }
\end{figure}

As explained in \citet{VinkIXPE} the test statistic for the detection of a polarized signal is
\begin{equation}
  S_{ij}\equiv \frac{Q_{ij}^2}{{\rm Var}(Q_{ij})} + \frac{U_{ij}^2}{{\rm Var}(U_{ij})},
\end{equation}
which has  $\chi^2$ distribution with two degrees of freedom, and relies on the fact that $Q$ and $U$ are orthogonal quantities.
The null-hypothesis is that there is no polarization signal, which implies that the covariances are expected to be zero.
Similarly we can now calculate
\begin{equation}
  \tilde{S}_{ij}\equiv \frac{\tilde{Q}_{ij}^2}{{\rm Var}(\tilde{Q}_{ij})} + \frac{\tilde{U}_{ij}^2}{{\rm Var}(\tilde{U}_{ij})},
\end{equation}
based on the smoothed $\tilde{Q}$ and $\tilde{U}$ maps, and their variances. In practice, the results are more stable against different pixel sizes
and choice of pixel centers. However, the map needs to be used with caution: neighboring pixel values are no longer independent.
Rather one can use it to indicate values of highly significant polarization, as it was used by \citet{FerrazzoliIXPE,Ferrazzoli2024}.

Figure~\ref{fig:smoothed} shows the result of the smoothing procedure with a Gaussian kernel with $\sigma=58^{\prime\prime}.3$.
The smoothing was done using a direct convolution, with spatial coordinates outside the map being assumed to be zero.
We only considered smoothed pixel values with a radius of  5$^{\prime}$ from the center of the map.
The highest polarization significance corresponds to $\chi^2_2=21.8$, corresponding to a $4.3\sigma$ significance.
The map contains 1128 data pixels. The Gaussian smoothing leads to a reduction in the number of resolution elements by a factor of 4, resulting in 282 independent $\chi^{2}_{2}$ values, based on the fact that the kernel size is twice as large as the IXPE angular resolution. Taking into account the 282 independent trials, the post-trial significance is close to 99.5\%.

\subsection{Spectro-polarimetric analysis}
A spectro-polarimetric analysis was conducted using the HEASoft package (version 6.33). 
The Level-2 data were reprocessed to eliminate high count rates unassociated with the source (see Section 2 for a detailed description). 
The source and background regions (as denoted in Figure \ref{plot_stokesI}) were first filtered using the HEASoft FTOOLS command \texttt{xselect}. 
The parameter \texttt{stokes=NEFF} was then set, and the weighted Stokes \(I\), \(Q\), and \(U\) spectra were extracted from the three DUs \citep[][]{DiMarco_2022}. 
The response files for the \(I\), \(Q\), and \(U\) spectra were generated using the IXPE mission-specific command \texttt{ixpecalcarf} with the cleaned Level-2 event list of each DU and the corresponding attitude files from housekeeping. The spectra were regrouped using the \texttt{ftgrouppha} command: \texttt{grouptype=min} and \texttt{groupscale}=500 for $I$ spectra, and \texttt{grouptype=constant} and \texttt{groupscale}=5 for $Q$ and $U$ spectra.

XSPEC (version 12.14.0) was used for fitting and plotting. 
All Stokes \(I\), \(Q\), and \(U\) spectra (background-subtracted) from all three DUs were simultaneously fitted using a model that consists of absorption, a simple power-law, and constant polarization degree and angle. 
To account for different flux calibrations of each DU, a cross-normalization constant factor, \texttt{const}, was also added. 
The model used in XSPEC is described as:
\[
\texttt{const}*\texttt{tbabs}(\texttt{polconst}*\texttt{powerlaw})
\]
The \texttt{tbabs} model \citep[][]{Wilms2000} was used to account for absorption along the line of sight toward the source.
Since IXPE spectra are most reliable between 2--8 keV, constraining the absorption using only photons above 2 keV is challenging. 
Therefore, the hydrogen column density \(N_H\) was fixed to \(0.4 \times 10^{22} \, \text{cm}^{-2}\), which was obtained from fitting a spectrum with an absorbed simple power-law model in the same region of a Chandra observation (ObsID: 9123) in the band from 0.8 to 4 keV.
Given that the IXPE spectrum is dominated by background above 4 keV, the analysis was limited to the 2--4 keV energy range. 

\begin{figure}
\includegraphics[width=0.45\textwidth]{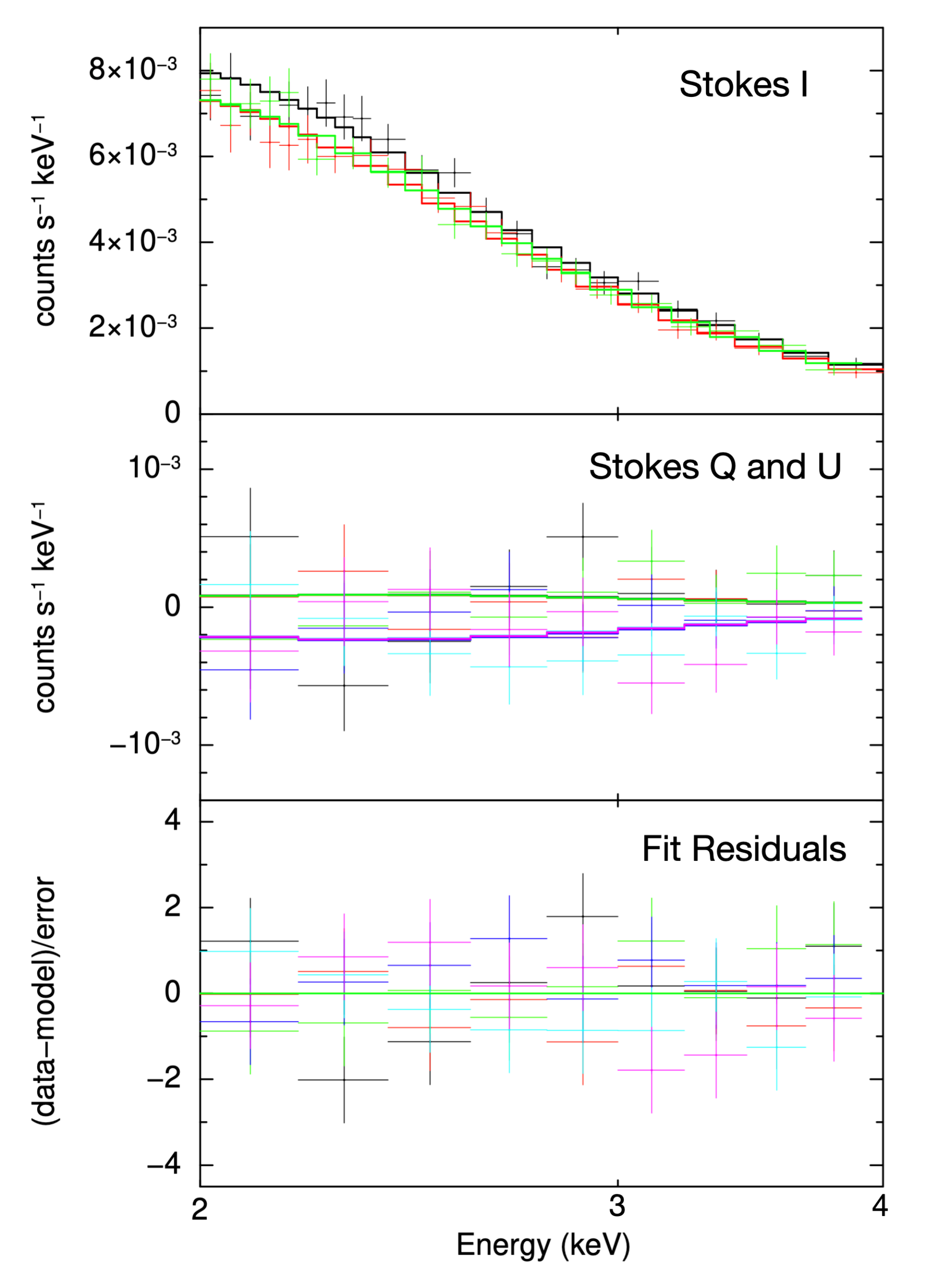}
\caption{Stokes \(I\), \(Q\), and \(U\) spectra of three DUs extracted from the source region with background subtraction (denoted in Figure \ref{plot_stokesI}). The three Stokes \(I\) spectra (DU1 in black, DU2 in red, DU3 in green) are shown in the upper panel. The six Stokes \(Q\) spectra (DU1 in black, DU2 in red, DU3 in green) and \(U\) spectra (DU1 in blue, DU2 in light blue, DU3 in magenta) are displayed in the middle panel. It appears that only two model lines (for $Q$ and $U$ spectra) are visible in this panel because the differences between the model lines for the three DUs are too small to be shown. The lower panel shows the fit residuals as \((\text{data} - \text{model}) / \text{error}\).}
\label{plot_spectra}
\end{figure}

\begin{table}[h!]
\centering
\caption{Best-fit spectro-polarimetric parameters of the source region from Vela Jr.}
\begin{tabular}{|c|c|c|}
\hline
Component & Parameter (unit) & Value \\
\hline
constant & factor DU1 & 1.000 (frozen) \\
 & factor DU2 & \(0.997_{-0.023}^{+0.024}\) \\
 & factor DU3 & \(1.012_{-0.023}^{+0.023}\) \\
\hline
TBabs & $N_\mathrm{H}$ (\(10^{22}\) cm\(^{-2}\)) & 0.400 (frozen) \\
polconst & A & \(0.175_{-0.047}^{+0.047}\) \\
polconst & \(\psi\) (deg) & \(-34.5_{-7.9}^{+7.9}\) $(145.5_{-7.9}^{+7.9})$  \\
powerlaw & \(\Gamma\) & \(2.41_{-0.06}^{+0.06}\) \\
\hline
\end{tabular}
\tablefoot{The hydrogen column density, $N_\mathrm{H}$, is fixed to the value obtained from the Chandra observation. Parameter \(A\) denotes PD, and \(\psi\) represents PA. The reduced chi-square value is \(\chi^2/\text{d.o.f.} = 70.57/112\). All values are quoted at a 68.3\% confidence level.}
\end{table}

The Stokes \(I\), \(Q\), and \(U\) spectra are shown in Figure \ref{plot_spectra} with the best-fit model and fit residuals. 
The spectrum is well described by an absorbed power-law model with a photon index \(\Gamma \approx 2.41\) in the 2--4 keV range. 
The total unabsorbed flux in the 2--8 keV range is approximately $(3.30^{+0.10}_{-0.09}) \times 10^{-12}$ ergs cm$^{-2}$ s$^{-1}$ at the 68.3\% confidence level.
PD\ \ $\approx17.5\%\pm 4.7\%$ and PA\ \ $\approx-34\fdg5 \pm 7\fdg9$ (equivalent to PA\ \ $=145\fdg5\pm7\fdg9$) were obtained from the polarization model \texttt{polconst}.
The best-fit parameters derived using HEASoft tools are listed in Table 5.

\section{Discussion}
\label{section5}

\begin{table*}
\centering
\caption{List of factors that may be responsible for radial polarization in RX J1713-3946 and Vela Jr. as discussed in Section \ref{section5}.}
\begin{tabular}{||c | c | c ||}
 \hline
 Parameter & Tangential-polarization mode & Radial-polarization mode \\
  & (in the radio and X-ray bands) & (RX J1713-3946 \& Vela Jr.) \\
 \hline
 Bohm factor, $\eta$ & high (2--15) if measured & close to 1 \\
 high TeV $\gamma$-ray luminosity & only true for Cas A & yes \\
 shock obliquity & likely parallel in SN 1006 & likely perpendicular \\
 age $>$1500 yr & only true for N132D \& Puppis A & yes \\
 dominance of synchrotron X-rays & only true for SNR G1.9+0.3 & yes \\
 low ISM density & only true for SN 1006 & yes \\
\hline
\end{tabular}
\label{TableFactors}
\end{table*}

There are many similarities between Vela Jr. and RX J1713.7-3946: (i) X-ray emission from both these SNRs are dominated by the synchrotron process (with a near absence of thermal X-ray emission); (ii) these two SNRs are strong sources of TeV $\gamma$-ray emission \citep[][]{Komin2005, Berge2006}; (iii) the Bohm factors measured in these remnants are smaller than in the first three SNRs observed with IXPE;
and last but not least, (iv) Vela Jr. and RX J1713.7-3946 have a radial polarization. The natural question to answer is whether the measured radial polarization can be related to some of the other mentioned facts. Table \ref{TableFactors} contains a summary of the main points discussed below.

The dominance of synchrotron X-rays is due to the expansion of Vela Jr. and RX J1713.7-3946 into a low density interstellar medium (ISM). These SNRs have not swept up a sufficient amount of matter to produce a comparable signal via free-free emission. However, the low density medium cannot be a decisive factor against a tangential polarization in SNRs; SN 1006 also expands into a low density medium, but has a tangential polarization \citep[][]{ZhouIXPE}. Additionally, the X-ray synchrotron-dominated SNRs include SNR G1.9+0.3 whose radio band polarization pattern is similar to that of SN 1006 \citep[][]{Luken2020}. Strong TeV $\gamma$-ray emission from Vela Jr. and RX J1713.7-3946 is most likely produced by inverse Compton scattering of cosmic-microwave-background photons by multi-TeV electrons \citep[e.g.,][]{Lee2013}. These two have high TeV $\gamma$-ray luminosity, but comparable to that of Cas A in which $\gamma$ rays have a hadronic origin \citep[e.g.,][]{Abeysekara2020}. Acceleration of electrons to multi-TeV energies is a requirement for production of both synchrotron X-rays and TeV $\gamma$ rays in Vela Jr. and RX J1713.7-3946. In diffusive shock acceleration, particles gradually gain energy by crossing the shock front forward and backward. Particles change their directions by being scattered by magnetic fields. If the Bohm factor, $\eta$, is 1, the particle mean free path takes the minimum value and the particles are accelerated most efficiently. The acceleration efficiency and the Bohm coefficient are interrelated. For example, the TeV $\gamma$-ray emission of SN 1006, which has a Bohm factor of $\simeq10$, is indeed less luminous \citep[][]{HESSSN1006} than that of Vela Jr. and RX J1713.7-3946. Since the Bohm factors of Vela Jr. (or RX J1713.7-3946) and SN 1006 are significantly different, this parameter can be related to their orthogonal orientations of polarization. At first glance, the fact that PD is as high as 13\%-16\% in Vela Jr. and RX J1713.7-3946 for strongly turbulent magnetic fields ($\eta=1$) may appear to be concerning. However, the Bohm factor measured by means of synchrotron X-ray spectral curvature \citep[][]{Zirakashvili2007} is for the direction along the normal to a shock front. Thus, the diffusive coefficient along the shock front can be larger than the Bohm diffusion coefficient and, in turn, the magnetic field along the shock front can be more regular \citep[][]{Casse2001}. In other words, PD in excess of 10\% for these two SNRs with $\eta=1$ may be measured -- when a polarization is radial -- in the case of anisotropic diffusion. Otherwise the PD values would be significantly lower.

Other factors may also be important in determining whether a polarization is radial or tangential. To explain the radial polarization of SNR G156.2+5.7 in the radio band, \citet[][]{Xu2007} suggested that the magnetic-field structure inferred from the observations of SNR G156.2+5.7 should reflect the magnetic field in the ambient ISM. The IMF is confined to the Galactic disk and is azimuthal \citep[][]{Han1994, Heiles1996}. Although the rims of Vela Jr. and RX J1713.7-3946 observed with IXPE are NW, the angular distance between these SNRs residing in the Galactic plane is $\sim80^{\circ}$. Thus, the NW rim of RX J1713.7-3946 expands along the normal to the Galactic equator, while the NW rim of Vela Jr. expands along the Galactic plane. Given the IMF configuration parallel to the Galactic plane, the polarization in the NW rim of RX J1713.7-3946 should be radial if it reflects the IMF. So it can be the case that the IMF significantly affects a polarization orientation in RX J1713.7-3946 and more evolved SNRs, such as SNR G156.2+5.7. Given the location of Vela Jr. relative to the Earth in the Galaxy, the IMF at the location of Vela Jr. is directed almost along the line of sight and is perpendicular to the shock normal for the rim of Vela Jr. The synchrotron mechanism allows one to measure only the magnetic-field orientation projected onto the plane of the sky. Therefore, the argument by \citet[][]{Xu2007} is not applicable to Vela Jr in the same way.
Meanwhile, the TeV $\gamma$-ray emission from the NW rims of Vela Jr. and RX J1713.7-3946 is strong, and the presence of multi-TeV electrons in these rims, likely corresponding to quasi-perpendicular regions of the shocks, has been established. It stands in contrast to the acknowledged fact that in regions where the shock is quasi-parallel (for which the average magnetic field direction upstream of the shock is close to the shock normal), electrons can be accelerated to multi-TeV energies \citep[][]{Park2015}, but in regions where the shock is quasi-perpendicular, electron acceleration occurs up to smaller energies \citep[][]{Xu2020}. It should be noted that in regions where the shock is quasi-perpendicular, ions are not injected into diffusive shock acceleration and the magnetic field may be not effectively amplified \citep[][]{Xu2020}. Although the theory of diffusive shock acceleration has long been the standard for cosmic-ray acceleration at shocks, other mechanisms of acceleration at collisionless quasi-perpendicular shocks are possible \citep[for a review, see][]{Amano2022}. We note that, in addition to the large-scale IMF, the local structure of the magnetic field near the SNR may shape the magnetic-field morphology in the SNR shell.

The age of an SNR is another parameter that can be relevant for explaining a polarization dichotomy. Before the launch of IXPE, radio observations revealed a tangential polarization in SNRs with ages less than 4000 years, but a radial polarization in older SNRs. In addition to Cas A, Tycho's SNR, and SN 1006, other SNRs with a tangential polarization are SN 1987A \citep[][]{Zanardo2018}, SNR G1.9+0.3 \citep[][]{Luken2020}, LMC N132D \citep[][]{Dickel1995}, and Puppis A \citep[][]{ Milne1993}. The list of older SNRs with a radial polarization includes SNR G182.4+4.3 \citep[][]{Kothes1998}, CTB 1 \citep[][]{Furst2004}, and SNR G156.2+5.7 \citep[][]{Xu2007}.  It is noteworthy that, in the radio band -- aside from the mentioned case of SNR 1E 0102.2-7219 -- the recent MeerKAT observations of two other young SNRs G4.8+6.2 and G7.7-3.7 revealed a radial polarization
\citep[][]{Cotton2024}.
It is accepted that a tangential polarization is characteristic for young SNRs in which the ejected material dominates the SNR dynamics. \citet[][]{Kothes1998} suggested that the radial polarization of SNR G182.2+4.3 indicates that dynamics of this SNR is dominated by the blast wave with the swept-up mass much larger than the ejecta mass. Thus, it is of importance to check if this mass ratio, $M_{\mathrm{swept-up}}/M_{\mathrm{ejecta}}$, is a factor determining a polarization structure.
RX J1713-3946 and Vela Jr. are young SNRs and their ages are $\simeq1600$ years and $\simeq3000$ years, respectively. These two SNRs did not show any evidence of thermal X-ray emission from highly ionized gas. This indicates that they have not yet swept up a significant amount of mass. This fact is in contrast to the measured radial polarization. Despite that LMC N132D is older than RX J1713-3946 and Puppis A is older than Vela Jr. and that both LMC N132D and Puppis A did show strong thermal X-ray emission from highly ionized gas, a tangential polarization was detected in LMC N132D and Puppis A. The IXPE polarization measurements in Vela Jr. and RX J1713-3946 showed that the trend -- the higher the mass ratio, $M_{\mathrm{swept-up}}/M_{\mathrm{ejecta}}$, is, the closer to radial the polarization is -- is not universal.

Finally, it is important to note that  the regularity degree of a magnetic field defines the ratio of transverse-to-parallel cosmic-ray diffusion coefficients. If the energy density in a regular magnetic field exceeds that in a turbulent magnetic field, then the parallel cosmic-ray diffusion coefficient can be many times larger than the transverse one \citep[see Figure 6 in][]{Casse2001}. The PD value provides us in turn with a measure of the ratio of regular-to-turbulent magnetic-field energy densities, or of the level of magnetic-field anisotropy \citep[][]{Korchakov1962, Bandiera2016, Bandiera2024}. In the case of a regular magnetic field, the maximum PD value possible for the X-ray spectrum with a photon index of $\simeq$2.4, corresponding to the NW rim of Vela Jr., is 78\% \citep[][]{Ginzburg1964}. The average PD value derived from the IXPE observations of this rim is $16.4\%\pm5.2\%$ after background subtraction. Meanwhile, the two maximal values of the observed PD for the pixels, that are bright in Stokes $I$, are $34.3\%\pm9.7\%$ and $32.5\%\pm9.0\%$ (Table \ref{Table3}) corresponding to the PD values of $85.2\%\pm30.1\%$ and $66.4\%\pm22.6\%$, respectively, after background subtraction. The latter two values are compatible with the maximum possible value. This compatibility may open up a new avenue for a further study of polarized X-rays from the western border of the NW rim, along which these two pixels are located, with future observations of Vela Jr.

\section{Conclusions}

The results of IXPE observations performed toward Cas A, Tycho's SNR, and SN 1006 showed that both the radial orientation of a magnetic field and the degree of magnetic-field regularity derived from polarized X-rays are comparable to those derived in the radio band. This relation indicates that the processes determining magnetic-field structures in these remnants act in very proximity to shock fronts. This is because X-rays, which are emitted by short-lived multi-TeV electrons, sample magnetic fields confined closer to the acceleration sites than polarized radio waves, which are emitted by long-lived GeV electrons.

The short lifetimes of multi-TeV electrons lead to a spectral curvature in X-ray spectra of young SNRs. The X-ray spectral curvature allows one to estimate the effectiveness of a magnetic turbulence in diffusing electrons across shock fronts, which is described by the Bohm factor. The smaller the Bohm factor, the higher the effectiveness of particle diffusion by a magnetic turbulence (or of particle acceleration).
The measured values of the Bohm factor for Cas A, Tycho's SNR, and SN 1006 are larger those in RX J1713.7-3946 and Vela Jr. For the latter two, the values of the Bohm factor are close to the smallest possible value, that is 1. The previous IXPE observations of RX J1713.7-3946 discovered a tangential magnetic-field orientation. This paper reports the results of IXPE observations of Vela Jr, the SNR which is similar to RX J1713.7-3946 in many ways. These results suggest the tangential orientation of a magnetic field in the NW rim of Vela Jr, making this SNR the second one with this magnetic-field orientation among the SNRs observed by IXPE. This field orientation along with the degree of polarization measured in the NW rim of Vela Jr., $PD=16.4\%\pm4.5\%$, similar to that for RX J1713.7-3946, $PD=13.0\%\pm3.5\%$, is indicative that the same process(es) leading to a tangential magnetic-field structure in both these SNRs.

The dichotomy in polarization between young and middle-aged SNRs was noticed in the radio band. It advocates radial magnetic fields in ejecta-dominated SNRs and tangential magnetic fields in middle-aged SNRs. If the relation between polarization properties in the radio and X-ray bands holds for the SNRs previously studied in the radio and X-ray bands, it allows a systematic study of polarization properties in SNRs. The results for RX J1713.7-3946 and Vela Jr. provide an important test-bed. These two SNRs have X-ray spectra dominated by synchrotron emission and do not show any thermal free-free X-ray emission. This means that they have not yet swept up a significant amount of mass and are dynamically young. Therefore, the  tangential orientation of magnetic fields in these two SNRs suggests that the evolutionary model in the context of the dichotomy in polarization does not spread to these two. The evolutionary model must be superseded by a theory also describing the magnetic-field structure in RX J1713.7-3946 and Vela Jr.

\begin{acknowledgements}
The Imaging X-ray Polarimetry Explorer (IXPE) is a joint US and Italian mission. The US contribution is supported by the National Aeronautics and Space Administration (NASA) and led and managed by its Marshall Space Flight Center (MSFC), with industry partner Ball Aerospace (contract NNM15AA18C). The Italian contribution is supported by the Italian Space Agency (Agenzia Spaziale Italiana, ASI) through contract ASI-OHBI-2022-13-I.0, agreements ASI-INAF-2022-19-HH.0 and ASI-INFN-2017.13-H0, and its Space Science Data Center (SSDC) with agreements ASI-INAF-2022-14-HH.0 and ASI-INFN 2021-43-HH.0, and by the Istituto Nazionale di Astrofisica (INAF) and the Istituto Nazionale di Fisica Nucleare (INFN) in Italy.
This research used data products provided by the IXPE Team (MSFC, SSDC, INAF, and INFN) and distributed with additional software tools by the High-Energy Astrophysics Science Archive Research Center (HEASARC) at NASA Goddard Space Flight Center (GSFC).
D.P. acknowledges support by the European Research Council, ERC Starting grant ``Mapping Highly-Energetic Messengers throughout the Universe'', under contract no. 949555.
R.F., E.Co., A.D.M., P.So., S.F., F.L.M., F.Mu. are partially supported by MAECI with grant CN24GR08 ``GRBAXP: Guangxi-Rome Bilateral Agreement for X-ray Polarimetry in Astrophysics''. N.B. was supported by the INAF MiniGrant ``PWNnumpol—Numerical Studies of Pulsar Wind Nebulae in the Light of IXPE''.
\end{acknowledgements}

\bibliographystyle{aa}

\begin{thebibliography}{77}
\expandafter\ifx\csname natexlab\endcsname\relax\def\natexlab#1{#1}\fi

\bibitem[{{Abeysekara} {et~al.}(2020){Abeysekara}, {Archer}, {Benbow}, {Bird},
  {Brose}, {Buchovecky}, {Buckley}, {Chromey}, {Cui}, {Daniel}, {Das},
  {Dwarkadas}, {Falcone}, {Feng}, {Finley}, {Fortson}, {Gent}, {Gillanders},
  {Giuri}, {Gueta}, {Hanna}, {Hassan}, {Hervet}, {Holder}, {Hughes},
  {Humensky}, {Kaaret}, {Kar}, {Kelley-Hoskins}, {Kertzman}, {Kieda}, {Krause},
  {Krennrich}, {Kumar}, {Lang}, {Maier}, {Moriarty}, {Mukherjee},
  {Nievas-Rosillo}, {O'Brien}, {Ong}, {Park}, {Petrashyk}, {Pfrang}, {Pohl},
  {Pueschel}, {Quinn}, {Ragan}, {Reynolds}, {Richards}, {Roache}, {Sadeh},
  {Santander}, {Sembroski}, {Shahinyan}, {Sushch}, {Weinstein}, {Wilcox},
  {Wilhelm}, {Williams}, {Williamson}, {Zitzer}, \& {Ghiotto}}]{Abeysekara2020}
{Abeysekara}, A.~U., {Archer}, A., {Benbow}, W., {et~al.} 2020, \apj, 894, 51

\bibitem[{{Acero} {et~al.}(2010){Acero}, {Aharonian}, {Akhperjanian}, {Anton},
  {Barres de Almeida}, {Bazer-Bachi}, \& {Becherini}}]{HESSSN1006}
{Acero}, F., {Aharonian}, F., {Akhperjanian}, A.~G., {et~al.} 2010, \aap, 516,
  A62

\bibitem[{{Aharonian} {et~al.}(2005){Aharonian}, {Akhperjanian}, {Bazer-Bachi},
  {Beilicke}, {Benbow}, {Berge}, \& {Bernl{\"o}hr}}]{Komin2005}
{Aharonian}, F., {Akhperjanian}, A.~G., {Bazer-Bachi}, A.~R., {et~al.} 2005,
  \aap, 437, L7

\bibitem[{{Aharonian} {et~al.}(2006){Aharonian}, {Akhperjanian}, {Bazer-Bachi},
  {Beilicke}, {Benbow}, {Berge}, \& {Bernl{\"o}hr}}]{Berge2006}
{Aharonian}, F., {Akhperjanian}, A.~G., {Bazer-Bachi}, A.~R., {et~al.} 2006,
  \aap, 449, 223

\bibitem[{{Allen} {et~al.}(2015){Allen}, {Chow}, {DeLaney}, {Filipovi{\'c}},
  {Houck}, {Pannuti}, \& {Stage}}]{Allen2015}
{Allen}, G.~E., {Chow}, K., {DeLaney}, T., {et~al.} 2015, \apj, 798, 82

\bibitem[{{Alsaberi} {et~al.}(2024){Alsaberi}, {Filipovi{\'c}}, {Dai}, {Sano},
  {Kothes}, {Payne}, {Bozzetto}, {Brose}, {Collischon}, {Crawford}, {Haberl},
  {Hill}, {Kavanagh}, {Knies}, {Leahy}, {Macgregor}, {Maggi}, {Maitra},
  {Manojlovi{\'c}}, {Mart{\'\i}n}, {Matthew}, {Ralph}, {Rowell}, {Ruiter},
  {Sasaki}, {Seitenzahl}, {Tokuda}, {Tothill}, {Uro{\v{s}}evi{\'c}}, {van
  Loon}, {Velovi{\'c}}, \& {Vogt}}]{Alsaberi2024}
{Alsaberi}, R. Z.~E., {Filipovi{\'c}}, M.~D., {Dai}, S., {et~al.} 2024, \mnras,
  527, 1444

\bibitem[{{Amano} {et~al.}(2022){Amano}, {Matsumoto}, {Bohdan}, {Kobzar},
  {Matsukiyo}, {Oka}, {Niemiec}, {Pohl}, \& {Hoshino}}]{Amano2022}
{Amano}, T., {Matsumoto}, Y., {Bohdan}, A., {et~al.} 2022, Reviews of Modern
  Plasma Physics, 6, 29

\bibitem[{{Aschenbach}(1998)}]{Aschenbach1998}
{Aschenbach}, B. 1998, \nat, 396, 141

\bibitem[{{Baldini} {et~al.}(2021){Baldini}, {Barbanera}, {Bellazzini},
  {Bonino}, {Borotto}, {Brez}, \& {Caporale}}]{Baldini2021}
{Baldini}, L., {Barbanera}, M., {Bellazzini}, R., {et~al.} 2021, Astroparticle
  Physics, 133, 102628

\bibitem[{{Baldini} {et~al.}(2022){Baldini}, {Bucciantini}, {Lalla}, {Ehlert},
  {Manfreda}, {Negro}, {Omodei}, {Pesce-Rollins}, {Sgr{\`o}}, \&
  {Silvestri}}]{Baldini2022}
{Baldini}, L., {Bucciantini}, N., {Lalla}, N.~D., {et~al.} 2022, SoftwareX, 19,
  101194

\bibitem[{{Bamba} {et~al.}(2005){Bamba}, {Yamazaki}, \& {Hiraga}}]{Bamba2005}
{Bamba}, A., {Yamazaki}, R., \& {Hiraga}, J.~S. 2005, \apj, 632, 294

\bibitem[{{Bandiera} \& {Petruk}(2016)}]{Bandiera2016}
{Bandiera}, R. \& {Petruk}, O. 2016, \mnras, 459, 178

\bibitem[{{Bandiera} \& {Petruk}(2024)}]{Bandiera2024}
{Bandiera}, R. \& {Petruk}, O. 2024, \aap, 689, A137

\bibitem[{{Bell}(1978)}]{Bell1978}
{Bell}, A.~R. 1978, \mnras, 182, 147

\bibitem[{{Bell}(2004)}]{Bell2004}
{Bell}, A.~R. 2004, \mnras, 353, 550

\bibitem[{{Bell} \& {Lucek}(2001)}]{Bell2001}
{Bell}, A.~R. \& {Lucek}, S.~G. 2001, \mnras, 321, 433

\bibitem[{{Bellazzini} {et~al.}(2007){Bellazzini}, {Spandre}, {Minuti},
  {Baldini}, {Brez}, {Latronico}, {Omodei}, {Razzano}, {Massai},
  {Pesce-Rollins}, {Sgr{\'o}}, {Costa}, {Soffitta}, {Sipila}, \&
  {Lempinen}}]{Bellazzini2007}
{Bellazzini}, R., {Spandre}, G., {Minuti}, M., {et~al.} 2007, Nuclear
  Instruments and Methods in Physics Research A, 579, 853

\bibitem[{{Berezinskii} {et~al.}(1990){Berezinskii}, {Bulanov}, {Dogiel}, \&
  {Ptuskin}}]{Berezinskii1990}
{Berezinskii}, V.~S., {Bulanov}, S.~V., {Dogiel}, V.~A., \& {Ptuskin}, V.~S.
  1990, {Astrophysics of cosmic rays}

\bibitem[{{Blandford} \& {Ostriker}(1978)}]{Blandford1978}
{Blandford}, R.~D. \& {Ostriker}, J.~P. 1978, \apjl, 221, L29

\bibitem[{{Bucciantini} {et~al.}(2023){Bucciantini}, {Di Lalla}, {Romani},
  {Silvestri}, {Negro}, {Baldini}, {Tennant}, \& {Manfreda}}]{Bucciantini2023}
{Bucciantini}, N., {Di Lalla}, N., {Romani}, R.~W.~R., {et~al.} 2023, \aap,
  672, A66

\bibitem[{{Camilloni} {et~al.}(2023){Camilloni}, {Becker}, {Predehl},
  {Dennerl}, {Freyberg}, {Mayer}, \& {Sasaki}}]{Camilloni2023}
{Camilloni}, F., {Becker}, W., {Predehl}, P., {et~al.} 2023, \aap, 673, A45

\bibitem[{{Casse} {et~al.}(2001){Casse}, {Lemoine}, \& {Pelletier}}]{Casse2001}
{Casse}, F., {Lemoine}, M., \& {Pelletier}, G. 2001, \prd, 65, 023002

\bibitem[{{Churazov} {et~al.}(2024){Churazov}, {Khabibullin}, {Barnouin},
  {Bucciantini}, {Costa}, {Di Gesu}, {Di Marco}, {Ferrazzoli}, {Forman},
  {Kaaret}, {Kim}, {Kolodziejczak}, {Kraft}, {Marin}, {Matt}, {Negro},
  {Romani}, {Silvestri}, {Soffitta}, {Sunyaev}, {Svoboda}, {Vikhlinin},
  {Weisskopf}, {Xie}, {Agudo}, {Antonelli}, {Bachetti}, {Baldini},
  {Baumgartner}, {Bellazzini}, {Bianchi}, {Bongiorno}, {Bonino}, {Brez},
  {Capitanio}, {Castellano}, {Cavazzuti}, {Chen}, {Ciprini}, {De Rosa}, {Del
  Monte}, {Di Lalla}, {Donnarumma}, {Doroshenko}, {Dov{\v{c}}iak}, {Ehlert},
  {Enoto}, {Evangelista}, {Fabiani}, {Garc{\'\i}a}, {Gunji}, {Hayashida},
  {Heyl}, {Iwakiri}, {Jorstad}, {Karas}, {Kislat}, {Kitaguchi}, {Krawczynski},
  {La Monaca}, {Latronico}, {Liodakis}, {Maldera}, {Manfreda}, {Marinucci},
  {Marscher}, {Marshall}, {Massaro}, {Mitsuishi}, {Mizuno}, {Muleri}, {Ng},
  {O'Dell}, {Omodei}, {Oppedisano}, {Papitto}, {Pavlov}, {Peirson}, {Perri},
  {Pesce-Rollins}, {Petrucci}, {Pilia}, {Possenti}, {Poutanen}, {Puccetti},
  {Ramsey}, {Rankin}, {Ratheesh}, {Roberts}, {Sgr{\`o}}, {Slane}, {Spandre},
  {Swartz}, {Tamagawa}, {Tavecchio}, {Taverna}, {Tawara}, {Tennant}, {Thomas},
  {Tombesi}, {Trois}, {Tsygankov}, {Turolla}, {Vink}, {Wu}, \&
  {Zane}}]{Churazov2023}
{Churazov}, E., {Khabibullin}, I., {Barnouin}, T., {et~al.} 2024, \aap, 686,
  A14

\bibitem[{{Costa} {et~al.}(2001){Costa}, {Soffitta}, {Bellazzini}, {Brez},
  {Lumb}, \& {Spandre}}]{Costa2001}
{Costa}, E., {Soffitta}, P., {Bellazzini}, R., {et~al.} 2001, \nat, 411, 662

\bibitem[{{Cotton} {et~al.}(2024){Cotton}, {Kothes}, {Camilo}, {Chandra},
  {Buchner}, \& {Nyamai}}]{Cotton2024}
{Cotton}, W.~D., {Kothes}, R., {Camilo}, F., {et~al.} 2024, \apjs, 270, 21

\bibitem[{{DeLaney} \& {Rudnick}(2003)}]{DeLaney2003}
{DeLaney}, T. \& {Rudnick}, L. 2003, \apj, 589, 818

\bibitem[{{Di Marco} {et~al.}(2022){Di Marco}, {Costa}, {Muleri}, {Soffitta},
  {Fabiani}, {La Monaca}, {Rankin}, {Xie}, {Bachetti}, {Baldini},
  {Baumgartner}, {Bellazzini}, {Brez}, {Castellano}, {Del Monte}, {Di Lalla},
  {Ferrazzoli}, {Latronico}, {Maldera}, {Manfreda}, {O'Dell}, {Perri},
  {Pesce-Rollins}, {Puccetti}, {Ramsey}, {Ratheesh}, {Sgrò}, {Spandre},
  {Tennant}, {Tobia}, {Trois}, \& {Weisskopf}}]{DiMarco_2022}
{Di Marco}, A., {Costa}, E., {Muleri}, F., {et~al.} 2022, The Astronomical
  Journal, 163, 170

\bibitem[{{Di Marco} {et~al.}(2023){Di Marco}, {Soffitta}, {Costa},
  {Ferrazzoli}, {La Monaca}, {Rankin}, {Ratheesh}, {Xie}, {Baldini}, {Del
  Monte}, {Ehlert}, {Fabiani}, {Kim}, {Muleri}, {O'Dell}, {Ramsey}, {Rubini},
  {Sgr{\`o}}, {Silvestri}, {Tennant}, \& {Weisskopf}}]{diMarco2023}
{Di Marco}, A., {Soffitta}, P., {Costa}, E., {et~al.} 2023, \aj, 165, 143

\bibitem[{{Dickel} \& {Milne}(1995)}]{Dickel1995}
{Dickel}, J.~R. \& {Milne}, D.~K. 1995, \aj, 109, 200

\bibitem[{{Dinsmore} \& {Romani}(2024)}]{Dinsmore2024}
{Dinsmore}, J.~T. \& {Romani}, R.~W. 2024, \apj, 962, 183

\bibitem[{{Dubner} \& {Giacani}(2015)}]{Dubner2015}
{Dubner}, G. \& {Giacani}, E. 2015, \aapr, 23, 3

\bibitem[{{Duncan} \& {Green}(2000)}]{Duncan2000}
{Duncan}, A.~R. \& {Green}, D.~A. 2000, \aap, 364, 732

\bibitem[{{Ferrazzoli} {et~al.}(2024){Ferrazzoli}, {Prokhorov}, {Bucciantini},
  {Slane}, {Vink}, {Cardillo}, {Yang}, \& {Silvestri}}]{Ferrazzoli2024}
{Ferrazzoli}, R., {Prokhorov}, D., {Bucciantini}, N., {et~al.} 2024, \apjl,
  967, L38

\bibitem[{{Ferrazzoli} {et~al.}(2023){Ferrazzoli}, {Slane}, {Prokhorov},
  {Zhou}, {Vink}, {Bucciantini}, \& {Costa}}]{FerrazzoliIXPE}
{Ferrazzoli}, R., {Slane}, P., {Prokhorov}, D., {et~al.} 2023, \apj, 945, 52

\bibitem[{{F{\"u}rst} \& {Reich}(2004)}]{Furst2004}
{F{\"u}rst}, E. \& {Reich}, W. 2004, in The Magnetized Interstellar Medium, ed.
  B.~{Uyaniker}, W.~{Reich}, \& R.~{Wielebinski}, 141--146

\bibitem[{{Ginzburg} \& {Syrovatskii}(1964)}]{Ginzburg1964}
{Ginzburg}, V.~L. \& {Syrovatskii}, S.~I. 1964, {The Origin of Cosmic Rays}

\bibitem[{{Gull}(1973)}]{Gull1973}
{Gull}, S.~F. 1973, \mnras, 161, 47

\bibitem[{{Han} \& {Qiao}(1994)}]{Han1994}
{Han}, J.~L. \& {Qiao}, G.~J. 1994, \aap, 288, 759

\bibitem[{{Heiles}(1996)}]{Heiles1996}
{Heiles}, C. 1996, \apj, 462, 316

\bibitem[{{Helder} {et~al.}(2012){Helder}, {Vink}, {Bykov}, {Ohira}, {Raymond},
  \& {Terrier}}]{Helder2012}
{Helder}, E.~A., {Vink}, J., {Bykov}, A.~M., {et~al.} 2012, \ssr, 173, 369

\bibitem[{{Inoue} {et~al.}(2013){Inoue}, {Shimoda}, {Ohira}, \&
  {Yamazaki}}]{Inoue2013}
{Inoue}, T., {Shimoda}, J., {Ohira}, Y., \& {Yamazaki}, R. 2013, \apjl, 772,
  L20

\bibitem[{{Jun} \& {Norman}(1996{\natexlab{a}})}]{JN962}
{Jun}, B.-I. \& {Norman}, M.~L. 1996{\natexlab{a}}, \apj, 472, 245

\bibitem[{{Jun} \& {Norman}(1996{\natexlab{b}})}]{JN961}
{Jun}, B.-I. \& {Norman}, M.~L. 1996{\natexlab{b}}, \apj, 465, 800

\bibitem[{{Kaaret} {et~al.}(2024){Kaaret}, {Ferrazzoli}, {Silvestri}, {Negro},
  {Manfreda}, {Wu}, {Costa}, {Soffitta}, {Safi-Harb}, {Poutanen}, {Veledina},
  {Di Marco}, {Slane}, {Bianchi}, {Ingram}, {Romani}, {Cibrario}, {Mac Intyre},
  {Mikus̆incov{\'a}}, {Ratheesh}, {Steiner}, {Svoboda}, {Tugliani}, {Agudo},
  {Antonelli}, {Bachetti}, {Baldini}, {Baumgartner}, {Bellazzini}, {Bongiorno},
  {Bonino}, {Brez}, {Bucciantini}, {Capitanio}, {Castellano}, {Cavazzuti},
  {Chen}, {Ciprini}, {De Rosa}, {Del Monte}, {Di Gesu}, {Di Lalla},
  {Donnarumma}, {Doroshenko}, {Dov{\v{c}}iak}, {Ehlert}, {Enoto},
  {Evangelista}, {Fabiani}, {Garc{\'\i}a}, {Gunji}, {Hayashida}, {Heyl},
  {Iwakiri}, {Jorstad}, {Karas}, {Kislat}, {Kitaguchi}, {Kolodziejczak},
  {Krawczynski}, {La Monaca}, {Latronico}, {Liodakis}, {Maldera}, {Marin},
  {Marinucci}, {Marscher}, {Marshall}, {Massaro}, {Matt}, {Mitsuishi},
  {Mizuno}, {Muleri}, {Ng}, {O'Dell}, {Omodei}, {Oppedisano}, {Papitto},
  {Pavlov}, {Peirson}, {Perri}, {Pesce-Rollins}, {Petrucci}, {Pilia},
  {Possenti}, {Puccetti}, {Ramsey}, {Rankin}, {Roberts}, {Sgr{\`o}}, {Spandre},
  {Swartz}, {Tamagawa}, {Tavecchio}, {Taverna}, {Tawara}, {Tennant}, {Thomas},
  {Tombesi}, {Trois}, {Tsygankov}, {Turolla}, {Vink}, {Weisskopf}, {Xie}, \&
  {Zane}}]{Kaaret2024}
{Kaaret}, P., {Ferrazzoli}, R., {Silvestri}, S., {et~al.} 2024, \apjl, 961, L12

\bibitem[{{Katsuda} {et~al.}(2008){Katsuda}, {Tsunemi}, \&
  {Mori}}]{Katsuda2008}
{Katsuda}, S., {Tsunemi}, H., \& {Mori}, K. 2008, \apjl, 678, L35

\bibitem[{{Korchakov} \& {Syrovatskii}(1962)}]{Korchakov1962}
{Korchakov}, A.~A. \& {Syrovatskii}, S.~I. 1962, \sovast, 5, 678

\bibitem[{{Kothes} {et~al.}(1998){Kothes}, {Furst}, \& {Reich}}]{Kothes1998}
{Kothes}, R., {Furst}, E., \& {Reich}, W. 1998, \aap, 331, 661

\bibitem[{{Koyama} {et~al.}(1997){Koyama}, {Kinugasa}, {Matsuzaki},
  {Nishiuchi}, {Sugizaki}, {Torii}, {Yamauchi}, \& {Aschenbach}}]{Koyama1997}
{Koyama}, K., {Kinugasa}, K., {Matsuzaki}, K., {et~al.} 1997, \pasj, 49, L7

\bibitem[{{Koyama} {et~al.}(1995){Koyama}, {Petre}, {Gotthelf}, {Hwang},
  {Matsuura}, {Ozaki}, \& {Holt}}]{Koyama1995}
{Koyama}, K., {Petre}, R., {Gotthelf}, E.~V., {et~al.} 1995, \nat, 378, 255

\bibitem[{{Krymskii}(1977)}]{Krymskii1977}
{Krymskii}, G.~F. 1977, Akademiia Nauk SSSR Doklady, 234, 1306

\bibitem[{{Kundu} \& {Velusamy}(1971)}]{Kundu1971}
{Kundu}, M.~R. \& {Velusamy}, T. 1971, \apj, 163, 231

\bibitem[{{Lazendic} {et~al.}(2004){Lazendic}, {Slane}, {Gaensler}, {Reynolds},
  {Plucinsky}, \& {Hughes}}]{Lazendic2004}
{Lazendic}, J.~S., {Slane}, P.~O., {Gaensler}, B.~M., {et~al.} 2004, \apj, 602,
  271

\bibitem[{{Lee} {et~al.}(2013){Lee}, {Slane}, {Ellison}, {Nagataki}, \&
  {Patnaude}}]{Lee2013}
{Lee}, S.-H., {Slane}, P.~O., {Ellison}, D.~C., {Nagataki}, S., \& {Patnaude},
  D.~J. 2013, \apj, 767, 20

\bibitem[{{Luken} {et~al.}(2020){Luken}, {Filipovi{\'c}}, {Maxted}, {Kothes},
  {Norris}, {Allison}, {Blackwell}, {Braiding}, {Brose}, {Burton}, {De Horta},
  {Galvin}, {Harvey-Smith}, {Hurley-Walker}, {Leahy}, {Ralph}, {Roper},
  {Rowell}, {Sushch}, {Uro{\v{s}}evi{\'c}}, \& {Wong}}]{Luken2020}
{Luken}, K.~J., {Filipovi{\'c}}, M.~D., {Maxted}, N.~I., {et~al.} 2020, \mnras,
  492, 2606

\bibitem[{{Mayer} {et~al.}(2023){Mayer}, {Becker}, {Predehl}, \&
  {Sasaki}}]{Mayer2023}
{Mayer}, M. G.~F., {Becker}, W., {Predehl}, P., \& {Sasaki}, M. 2023, \aap,
  676, A68

\bibitem[{{Milne} {et~al.}(1993){Milne}, {Stewart}, \& {Haynes}}]{Milne1993}
{Milne}, D.~K., {Stewart}, R.~T., \& {Haynes}, R.~F. 1993, \mnras, 261, 366

\bibitem[{{Parizot} {et~al.}(2006){Parizot}, {Marcowith}, {Ballet}, \&
  {Gallant}}]{Parizot2006}
{Parizot}, E., {Marcowith}, A., {Ballet}, J., \& {Gallant}, Y.~A. 2006, \aap,
  453, 387

\bibitem[{{Park} {et~al.}(2015){Park}, {Caprioli}, \& {Spitkovsky}}]{Park2015}
{Park}, J., {Caprioli}, D., \& {Spitkovsky}, A. 2015, \prl, 114, 085003

\bibitem[{{Pavlov} {et~al.}(2001){Pavlov}, {Sanwal}, {K{\i}z{\i}ltan}, \&
  {Garmire}}]{Pavlov2001}
{Pavlov}, G.~G., {Sanwal}, D., {K{\i}z{\i}ltan}, B., \& {Garmire}, G.~P. 2001,
  \apjl, 559, L131

\bibitem[{{Ramsey} {et~al.}(2022){Ramsey}, {Bongiorno}, {Kolodziejczak},
  {Kilaru}, {Alexander}, {Baumgartner}, {Breeding}, {Elsner}, {Le Roy},
  {McCracken}, {Mitsuishi}, {O'Dell}, {Pavelitz}, {Ranganathan}, {Sanchez},
  {Speegle}, {Thomas}, {Weddendorf}, \& {Weisskopf}}]{Ramsey2022}
{Ramsey}, B.~D., {Bongiorno}, S.~D., {Kolodziejczak}, J.~J., {et~al.} 2022,
  Journal of Astronomical Telescopes, Instruments, and Systems, 8, 024003

\bibitem[{{Reynolds}(2008)}]{Reynolds2008}
{Reynolds}, S.~P. 2008, \araa, 46, 89

\bibitem[{{Reynoso} {et~al.}(2013){Reynoso}, {Hughes}, \&
  {Moffett}}]{Reynoso2013}
{Reynoso}, E.~M., {Hughes}, J.~P., \& {Moffett}, D.~A. 2013, \aj, 145, 104

\bibitem[{{Rosenberg}(1970)}]{Rosenberg1970}
{Rosenberg}, I. 1970, \mnras, 151, 109

\bibitem[{{Slane} {et~al.}(2001){Slane}, {Hughes}, {Edgar}, {Plucinsky},
  {Miyata}, {Tsunemi}, \& {Aschenbach}}]{Slane2001}
{Slane}, P., {Hughes}, J.~P., {Edgar}, R.~J., {et~al.} 2001, \apj, 548, 814

\bibitem[{{Tsuji} {et~al.}(2021){Tsuji}, {Uchiyama}, {Khangulyan}, \&
  {Aharonian}}]{Tsuji2021}
{Tsuji}, N., {Uchiyama}, Y., {Khangulyan}, D., \& {Aharonian}, F. 2021, \apj,
  907, 117

\bibitem[{{Vink}(2012)}]{Vink2012}
{Vink}, J. 2012, \aapr, 20, 49

\bibitem[{{Vink} \& {Laming}(2003)}]{Vink2003}
{Vink}, J. \& {Laming}, J.~M. 2003, \apj, 584, 758

\bibitem[{{Vink} {et~al.}(2022){Vink}, {Prokhorov}, {Ferrazzoli}, {Slane},
  {Zhou}, {Asakura}, \& {Baldini}}]{VinkIXPE}
{Vink}, J., {Prokhorov}, D., {Ferrazzoli}, R., {et~al.} 2022, \apj, 938, 40

\bibitem[{{V{\"o}lk} {et~al.}(2005){V{\"o}lk}, {Berezhko}, \&
  {Ksenofontov}}]{Voelk2005}
{V{\"o}lk}, H.~J., {Berezhko}, E.~G., \& {Ksenofontov}, L.~T. 2005, \aap, 433,
  229

\bibitem[{{Weisskopf} {et~al.}(2022){Weisskopf}, {Soffitta}, {Baldini},
  {Ramsey}, {O'Dell}, {Romani}, \& {Matt}}]{Weisskopf2022}
{Weisskopf}, M.~C., {Soffitta}, P., {Baldini}, L., {et~al.} 2022, Journal of
  Astronomical Telescopes, Instruments, and Systems, 8, 026002

\bibitem[{{West} {et~al.}(2017){West}, {Jaffe}, {Ferrand}, {Safi-Harb}, \&
  {Gaensler}}]{West2017}
{West}, J.~L., {Jaffe}, T., {Ferrand}, G., {Safi-Harb}, S., \& {Gaensler},
  B.~M. 2017, \apjl, 849, L22

\bibitem[{{Wilms} {et~al.}(2000){Wilms}, {Allen}, \& {McCray}}]{Wilms2000}
{Wilms}, J., {Allen}, A., \& {McCray}, R. 2000, \apj, 542, 914

\bibitem[{{Xu} {et~al.}(2007){Xu}, {Han}, {Sun}, {Reich}, {Xiao}, {Reich}, \&
  {Wielebinski}}]{Xu2007}
{Xu}, J.~W., {Han}, J.~L., {Sun}, X.~H., {et~al.} 2007, \aap, 470, 969

\bibitem[{{Xu} {et~al.}(2020){Xu}, {Spitkovsky}, \& {Caprioli}}]{Xu2020}
{Xu}, R., {Spitkovsky}, A., \& {Caprioli}, D. 2020, \apjl, 897, L41

\bibitem[{{Zanardo} {et~al.}(2018){Zanardo}, {Staveley-Smith}, {Gaensler},
  {Indebetouw}, {Ng}, {Matsuura}, \& {Tzioumis}}]{Zanardo2018}
{Zanardo}, G., {Staveley-Smith}, L., {Gaensler}, B.~M., {et~al.} 2018, \apjl,
  861, L9

\bibitem[{{Zhou} {et~al.}(2023){Zhou}, {Prokhorov}, {Ferrazzoli}, {Yang},
  {Slane}, {Vink}, \& {Silvestri}}]{ZhouIXPE}
{Zhou}, P., {Prokhorov}, D., {Ferrazzoli}, R., {et~al.} 2023, \apj, 957, 55

\bibitem[{{Zirakashvili} \& {Aharonian}(2007)}]{Zirakashvili2007}
{Zirakashvili}, V.~N. \& {Aharonian}, F. 2007, \aap, 465, 695

\end{thebibliography}

\end{document}